\documentclass[12pt,twoside,openright]{article}
\usepackage[utf8]{inputenc}
\usepackage[a4paper, width=150mm, top=25mm, bottom=25mm, bindingoffset=6mm]{geometry}
\usepackage{fancyhdr}
\usepackage{epsfig,epic,eepic}
\usepackage{hyperref} 
\usepackage{textcomp} 
\usepackage{url} 
\usepackage{longtable} 
\usepackage{mathrsfs} 
\usepackage{array} 
\usepackage{multirow} 
\usepackage{bigstrut} 
\usepackage{amssymb} 
\usepackage{amsmath} 
\usepackage{adjustbox}
\usepackage{graphicx}
\usepackage{lscape} 
\usepackage{notoccite} 
\usepackage{enumitem}
\usepackage{adjustbox}
\usepackage{pdfpages}
\newcolumntype{C}[1]{>{\centering\arraybackslash}m{#1}}

\usepackage{ragged2e}

\usepackage[nottoc]{tocbibind} 
\usepackage{chemmacros} 
\usepackage{textgreek} 
\usepackage{siunitx} 
\usepackage{enumitem}
\definecolor{Constructive}{HTML}{ff0000}
\definecolor{Strengthening}{HTML}{ffc020}
\definecolor{Weakening}{HTML}{00c000}
\definecolor{Destructive}{HTML}{0080ff}
\usepackage[style=numeric, sorting=none]{biblatex}
\usepackage[utf8]{inputenc}
\usepackage[T1]{fontenc}
\usepackage{lmodern}

\begin{document}

 \begin{titlepage}
\begin{center}
         \includegraphics[width=0.5\textwidth]{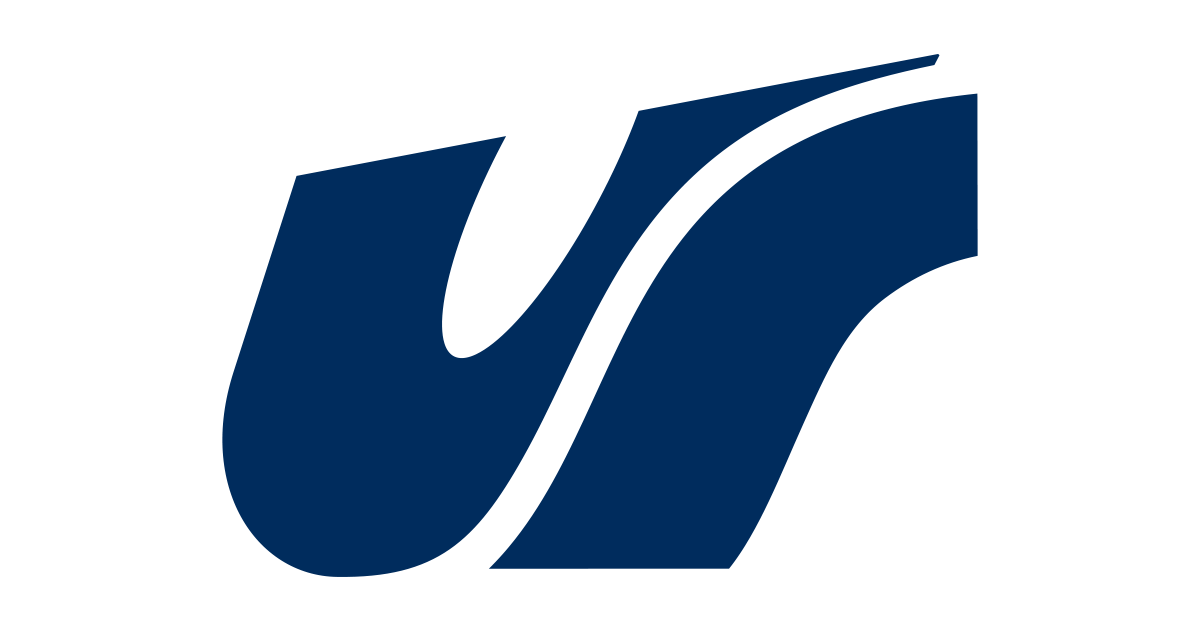} 
		        
		\vspace{0.5cm}

		\Large
		University of Silesia in Katowice		
		
		Faculty of Science and Technology       
		
		August Chełkowski Institute of Physics  
		\vspace{1cm}
		
		  \Large
        Doctoral Dissertation
         \vspace{2cm}

        \LARGE
        \textbf{Anomalous transport in periodic systems\\
         driven by active fluctuations}\\
        \vspace{0.5cm}
     
        \vspace{1.5cm}
     \Large   
        \textbf{Karol Białas} 
        
        \vfill
        
        \Large
            
        \vfill
            
        \Large 
        \textbf{Supervisor:} dr hab. Jakub Spiechowicz, prof. UŚ

        \vfill
            
        \large
        Chorzów, 2026
            
    \end{center}

\end{titlepage}
 \newpage
 
 \thispagestyle{empty}
 \
 
 \newpage
 
 \pagenumbering{roman}


    \vspace{1.5cm}
    \section*{Abstract}
    \phantomsection
    \addcontentsline{toc}{section}{Abstract}
    \vspace{0.5cm}
    
This dissertation investigates the transport of Brownian particles driven by active fluctuations in periodic potential landscapes. Unlike thermal noise, active fluctuations break detailed balance and violate the fluctuation-dissipation theorem. They model self-propulsion mechanisms and active-bath interactions in biological and synthetic micro-systems. By focusing on white Poisson shot noise with non-Gaussian amplitude statistics, this work systematically uncovers the mechanisms governing three major transport anomalies: giant transport enhancement, absolute negative mobility, and the ratchet effect.

Combining high-precision, GPU-accelerated Monte Carlo simulations with analytical modeling, this work demonstrates that active fluctuations adhering to a bidirectional skew-normal amplitude distribution enhance directed velocity compared to the free-particle by several orders of magnitude. This giant transport enhancement is fully quantified via a phenomenological jump-relaxation model that decouples particle displacement into instantaneous stochastic jumps and deterministic relaxation towards potential minimum. Extending this model to inertial dynamics reveals that finite particle mass $m$ acts as a dynamic control parameter capable of strengthening, weakening, constructively inducing, or completely suppressing transport enhancement depending on noise variance.

Furthermore, an ultra-minimal model for absolute negative mobility is established in a one-dimensional, purely overdamped system, showing that discontinuous shot noise drives transport opposite to an applied bias without requiring particle inertia, nonlinearity of spatially periodic potential, or explicit time-dependent periodic driving. Conversely, under an alternative scaling of the average stochastic force, an asymmetric distribution of fluctuation amplitudes generates net directed transport through a tilting ratchet effect in the zero-bias limit. Ultimately, these results provide theoretical insights into nonequilibrium statistical physics and offer design principles for synthetic nanomotors and microfluidic sorting devices, with direct relevance to intracellular transport such as kinesin and dynein movement along microtubules, and experimental realizations, e.g., via optical tweezers.


\newpage
    \vspace{1.5cm}
    \section*{Streszczenie}
    \phantomsection
    \vspace{0.5cm}

Niniejsza rozprawa doktorska jest poświęcona badaniu transportu cząstek Browna napędzanych fluktuacjami aktywnymi w potencjałach okresowych. W przeciwieństwie do szumu termicznego, fluktuacje aktywne łamią zasadę równowagi szczegółowej oraz naruszają twierdzenie fluktuacyjno-dyssypacyjne. Modelują one mechanizmy samonapędu i oddziaływania z zawiesiną materii aktywnej w mikroukładach biologicznych oraz syntetycznych. Skupiając się na białym szumie Poissona o niegaussowskiej statystyce amplitud, w pracy w sposób systematyczny opisano mechanizmy rządzące trzema głównymi anomaliami transportu: gigantycznym wzmocnieniem transportu, bezwzględną ujemną ruchliwością oraz efektem ratchetowym.

Łącząc precyzyjne symulacje Monte Carlo na procesorach graficznych (GPU) z modelowaniem analitycznym, wykazano, że fluktuacje aktywne opisane dwukierunkowym skośno-normalnym rozkładem amplitud zwiększają prędkość cząstki, w porównaniu do cząstki swobodnej nawet o kilka rzędów wielkości. Owo gigantyczne wzmocnienie transportu zostało w pełni opisane ilościowo za pomocą fenomenologicznego modelu skok-relaksacja, który rozdziela przemieszczenie cząstki na natychmiastowe skoki stochastyczne i deterministyczną relaksację do minimum potencjału. Uwzględnienie inercji pokazuje, że określona masa $m$ cząstki pełni funkcję dynamicznego parametru kontrolnego, zdolnego do wzmacniania, osłabiania, konstruktywnego indukowania lub całkowitego tłumienia wzmocnienia transportu w zależności od wariancji aktywnego szumu.

Ponadto sformułowano ultraminimalny model bezwzględnej ujemnej ruchliwości w jednowymiarowym, przetłumionym układzie, wykazując, że nieciągły szum Poissona wywołuje transport skierowany przeciwnie do przyłożonej zewnętrznej siły bez konieczności uwzględniania bezwładności cząstki, nieliniowego potencjału czy jawnie zależnego od czasu wymuszenia okresowego. Z kolei przy alternatywnym skalowaniu średniej siły stochastycznej asymetryczny rozkład amplitud fluktuacji wywołuje skierowany transport wypadkowy dzięki efektowi ratchetowemu w granicy zerowej siły.

Wyniki te dostarczają teoretycznego wglądu w nierównowagową fizykę statystyczną oraz oferują wskazówki do projektowania syntetycznych nanomotorów i mikroprzepływowych urządzeń sortujących, wykazując bezpośrednie odniesienie do transportu wewnątrzkomórkowego takiego jak ruch kinezyny i dyneiny wzdłuż mikrotubul, oraz możliwości realizacji eksperymentalnej, np. za pomocą szczypcy optycznych.

\newpage

    \vspace{1.5cm}
    {
    \section*{Acknowledgements} 
    \phantomsection 
    \addcontentsline{toc}{section}{Acknowledgements} 
    \vspace{0.5cm}
I would like to express my gratitude for my supervisor Jakub Spiechowicz for guidance during my whole studies at University of Silesia and for all the interesting and informative talks over the years. In particular I want to thank him for all the corrections and feedback during the preparation of this thesis. 

I would also like to thank Mateusz Wiśniewski for partnership over the course of our PhD. In addition, I would like to acknowledge my parents' continuous support. Last but not least, I would like to thank my uncle Władysław Borgieł for pointing me towards physics all these years ago.
}
\newpage

    \vspace{1.5cm}
    \section*{List of Publications}
    \phantomsection
    \addcontentsline{toc}{section}{List of Publications}
    \vspace{0.5cm}

This thesis serves as the introduction to the collection of the following articles about transport in periodic systems driven by active fluctuations
\begin{enumerate}[label=A\arabic*]
    \item K. Białas, J. Łuczka, J. Spiechowicz, Periodic potential can enormously boost free particle transport induced by active fluctuations, Physical Review E 107 (2), 024107 (2023)
    \item K. Białas, J. Spiechowicz, Mechanism for giant enhancement of transport induced by active fluctuations, Physical Review E 107 (6), 064120 (2023) \textit{(Editors' Suggestion)}
    \item K. Białas, J. Łuczka, J. Spiechowicz, Control of particle transport driven by active noise: strategy of amplification via a periodic potential, The European Physical Journal Special Topics 232 (20), 3191-3196 (2023)
    \item K. Białas, J. Spiechowicz Inertia-induced mechanism for giant enhancement of transport generated by active fluctuations, Chaos: An Interdisciplinary Journal of Nonlinear Science 35 (6) (2025)
    \item K. Białas, J. Spiechowicz, Giant enhancement of transport driven by active fluctuations: Impact of inertia, Physical Review E 112 (6), 064118 (2025)
    \item K. Białas, P. Hänggi, J. Spiechowicz, Active fluctuations drive inertia-free absolute negative mobility in soft and living matter, in press in Communications Physics (Nature group), arXiv preprint arXiv:2603.06070 (2026)
\end{enumerate}
Other articles by the author of this thesis about diffusion in analogous systems
\begin{enumerate}[label=B\arabic*]
    \item K. Białas, J. Łuczka, P. Hänggi, J. Spiechowicz, Colossal Brownian yet non-Gaussian diffusion induced by nonequilibrium noise, Physical Review E 102 (4), 042121 (2020)
    \item K, Białas, J. Spiechowicz, Colossal Brownian yet non-Gaussian diffusion in a periodic potential: Impact of nonequilibrium noise amplitude statistics, Chaos: An Interdisciplinary Journal of Nonlinear Science 31 (12) (2021)
\end{enumerate}

\newpage
\thispagestyle{empty}

\begin{center}
\LARGE
\hspace{0pt}
\vspace{0.5cm}
\vfill
GUIDEBOOK
\vfill
\hspace{0pt}
\end{center} 
\renewcommand*\contentsname{Table of Contents} 
 \tableofcontents 

\include{Lista}

\pagenumbering{arabic} 
\setcounter{page}{1} 

\section{Introduction}

Historically viewed as mere experimental artifacts or measurement errors, physical fluctuations have profoundly reshaped our understanding of the microscopic world, evolving from a macroscopic nuisance into a fundamental pillar of modern statistical physics. This evolution began with Robert Brown's discovery of the random jittering of pollen grains suspended in a fluid in 1827 \cite{brown1828xxvii}, which Einstein and Smoluchowski deciphered almost 80 years later \cite{einstein1905molekularkinetischen,von1906kinetischen}. 
In 1908, Perrin experimentally verified the results of Einstein and Smoluchowski \cite{perrin1909mouvement}, thereby definitively proving the atomic nature of matter. As electronics rapidly developed, Schottky discovered a new type of noise in 1918: shot noise, a current fluctuation caused by the random discharge of electrons from the cathode in a vacuum tube \cite{schottky1918spontane}. A decade later, Johnson demonstrated \cite{johnson1928thermal} and Nyquist theoretically explained \cite{nyquist1928thermal} that random voltage fluctuations arise from the thermal agitation of electrons in resistors. 
These seemingly disparate phenomena in fluids and circuits were ultimately unified in the 1950s by Callen, Welton, and Kubo via the Fluctuation-Dissipation Theorem \cite{callen1951irreversibility, kubo1966fluctuation}. This theorem established a universal relationship between a system's spontaneous fluctuations at equilibrium and its macroscopic response to external perturbations. In recent years, the study of nonequilibrium, nonthermal fluctuations represents an active area of ongoing research.

Within contemporary statistical physics, active matter has emerged as a major frontier. Active matter systems consume energy from their environment to self-propel, and they can be of both biological and artificial origin \cite{RevModPhys.85.1143,bechinger2016active,needleman2017active}. Examples of the former include motile bacteria, such as \textit{E. coli} \cite{gutierrez2018induced}, or spermatozoa \cite{tung2017fluid}. The latter can be physically realized, for instance, with chiral particles \cite{liebchen2022chiral}, Janus particles (characterized by two distinct ``faces'') \cite{das2015boundaries}, or rods \cite{vutukuri2016dynamic}. The fluctuations arising from these random self-propulsion mechanisms, alongside collisions with motile active agents, differ fundamentally from thermal fluctuations \cite{fodor2016far,dabelow2019irreversibility,o2022time}. Such athermal noise is termed \textit{active fluctuations}. Because active matter systems operate inherently far from equilibrium, the traditional hallmarks of equilibrium statistical mechanics, such as detailed balance \cite{fodor2016far,o2022time} and the fluctuation-dissipation theorem \cite{marconi2008fluctuation,dabelow2019irreversibility,maggi2017memory} do not apply to active fluctuations.

Biological systems can extract work from these nonequilibrium fluctuations. Kinesin and dynein, for example, are motor proteins that convert chemical energy from adenosine triphosphate (ATP) into mechanical motion \cite{gilbert1995pathway,ross2006processive}. They transport organelles and proteins within living cells along periodic tracks known as microtubules. A deeper understanding of these transport mechanisms could facilitate the development of highly efficient artificial nano- and microrobots driven by active fluctuations \cite{wang2012nano}. Such devices hold significant promise for biomedical applications, including targeted drug delivery. This potential serves as the primary motivation for this thesis: investigating transport phenomena in periodic systems driven by active fluctuations. In particular, I focus on the transport anomalies that emerge within these nonequilibrium settings.

The advent of modern experimental techniques, most notably optical tweezers \cite{ashkin1970acceleration,ashkin1986observation,neuman2004optical}, which garnered the 2018 Nobel Prize in Physics, has provided physicists with an unprecedented window into previously inaccessible microscopic dynamics. By employing highly focused laser beams, experimentalists can generate a harmonic potential well capable of trapping biological entities, such as cells, motile bacteria, and organelles \cite{norregaard2017manipulation,arbore2019probing,diekmann2016nanoscopy} as well as artificial objects, like polystyrene beads or nanoparticles \cite{juan2011plasmon}. Near the center of the trap, this potential is well-approximated as harmonic. The parabolic nature of the optical trap is particularly advantageous for precise measurements \cite{rohrbach2002trapping}. By measuring the displacement $x$ of a trapped object from the trap center, one can ascertain the exact restoring force acting upon it via Hooke's law, $F=-kx$, where $k$ is the stiffness of the optical tweezers. Notably, the previously elusive instantaneous velocity of Brownian particles has been successfully measured utilizing this technique \cite{li2010measurement,huang2011direct}.

However, passive measurements represent only a fraction of the capabilities offered by optical tweezers. By actively modulating the trap's center, experimentalists can exert precisely controlled forces on a trapped object \cite{kumar2018nanoscale,albay2018optical}. This technique allows for the creation of ``virtual'' potentials such as periodic energy landscapes or the application of tailored stochastic forces. Consequently, experimentalists can physically realize and investigate complex systems that were previously accessible only via numerical simulations, e.g., perfectly tunable active matter and active fluctuations.

\subsection{Synopsis}
This thesis provides a general introduction to the field of active fluctuations and an overview of the results published in the accompanying articles. 

In the following chapter, thermal fluctuations and their core characteristics are derived through a simple Gedankenexperiment. 

Chapter 3 contrasts thermal noise with active fluctuations, emphasizing their fundamental physical differences. This chapter also introduces primary stochastic processes used to model active systems: the Ornstein-Uhlenbeck process and white Poisson shot noise, the latter being the central focus of this thesis. Other models are also briefly described.

Chapter 4 covers the foundational theory of Brownian particle transport under an external force, with a particular emphasis on dynamics within periodic structures. 

Chapter 5 outlines key transport anomalies, namely giant transport enhancement, absolute negative mobility, and the ratchet effect.

Chapters 6 through 8 present the original research findings of this work. Chapter 6 details the counterintuitive giant enhancement of directed transport and in the first section introduces the phenomenological jump-relaxation model used to quantify this effect. 
The second section explores the role of inertia in this transport mechanism, discussing how inertial effects alter the system dynamics and the magnitude of the velocity enhancement. 

Chapter 7 focuses on the parameter regime in which absolute negative mobility can be detected, carefully delineating the necessary conditions for its emergence. 

Chapter 8 completes the analysis by presenting an alternative scaling for Poisson shot noise and establishing the conditions required for the emergence of the ratchet effect. Finally, the thesis concludes with a brief summary, closing remarks, and an outline of potential future research directions.

\section{Thermal Fluctuations}
\label{sec:thermaldynamics}

A particle immersed in a fluid undergoes countless collisions with the surrounding molecules. These collisions are an archetypal example of the thermal fluctuations occurring in any physical system at a nonzero temperature $T$. To determine the characteristics of these fluctuations, one can consider a simple \textit{Gedankenexperiment}, such as a Rayleigh piston, a moving wall in a narrow tube containing a dilute gas at thermal equilibrium on both sides. For a rigorous description of its dynamics, we refer the reader to van Kampen \cite{van1982diffusion,alkemade1963non,van1983stochastic}. 

At thermal equilibrium, the dynamics is reversible at the microscale. This property is dictated by detailed balance \cite{huang2009introduction}, which is a necessary but not sufficient condition for macroscopic equilibrium. For a system in state $A$ with probability $P(A)$, detailed balance requires that the transition to state $B$ with transition probability $W(A\to B)$ is strictly balanced by the reverse transition:
\begin{equation}
    W(A \to B)P(A) = W(B \to A)P(B).
\end{equation}
Consequently, thermal fluctuations are symmetric: a collision of amplitude $\theta$ affecting the particle is just as likely as one with an amplitude $-\theta$. Due to this micro-reversibility, no net entropy is produced by thermal fluctuations. If $\rho(\theta)$ denotes the probability density function of the fluctuation amplitudes, it follows that:
\begin{equation}
    \rho(\theta)=\rho(-\theta).
\end{equation}
Because this distribution is symmetric, the average fluctuation amplitude must vanish:
\begin{equation}
    \langle \theta \rangle = \int_{-\infty}^{\infty}\theta\rho(\theta)d\theta = 0.
\end{equation}

Rather than analyzing individual collisions, we model the effective thermal fluctuations as a stochastic process, $\Gamma(t)$. Because an equilibrium system remains in a steady state, it is time-translation invariant. Consequently, $\Gamma(t)$ must be a stationary process, meaning that its statistical properties, such as its mean $\langle\Gamma(t)\rangle$ and autocorrelation $\langle\Gamma(t)\Gamma(s)\rangle$, are independent of time $t$:
\begin{equation}
    \langle \Gamma(t)\rangle=\text{const},
\end{equation}
\begin{equation}
    \langle\Gamma(t)\Gamma(s)\rangle=R(t-s).
\end{equation}
Similarly, the average number of collisions per unit time is constant:
\begin{equation}
    \lim_{t\to\infty}\frac{\langle n_c(t)\rangle}{t} = \lambda_c.
\end{equation}

Because the duration of a single collision is much shorter than any other relevant timescale, collisions can be treated as instantaneous and mutually independent. Consequently, these thermal fluctuations can be modeled as white Poisson shot noise \cite{hanggi1978derivations}:
\begin{equation}
    \Gamma(t)=\sum_{i=1}^{n_c(t)}\theta_i\delta(t-t_i),
    \label{poisson_thermal}
\end{equation}
where $\theta_i$ is the collision amplitude and $t_i$ is the collision time; the sequence of collisions is governed by a Poisson counting process $n_c(t)$ with rate $\lambda_c$ \cite{feller1991introduction}. The probability of exactly $k$ collisions occurring in the time interval $[0,t]$ is:
\begin{equation}
    \text{Pr}\{n_c(t) = k\} = \frac{(\lambda_c t)^k}{k!} e^{-\lambda_c t}.
\end{equation}
This stochastic process has a zero mean and is delta-correlated:
\begin{equation}
    \langle \Gamma(t)\rangle = \lambda_c\langle\theta_i\rangle = 0,
\end{equation}
\begin{equation}
    \langle\Gamma(t)\Gamma(s)\rangle = 2D_T\delta(t-s),
\end{equation}
where $D_T=\lambda_c\langle \theta_i^2\rangle/2$ is the Poisson noise intensity \cite{spiechowicz2014brownian} and $\langle\theta_i^2\rangle$ is the second moment of the random amplitude. The delta-correlated (white) nature of $\Gamma(t)$ implies that the process has no memory. 

Physically, the average time between molecular collisions, $\tau_c = 1/\lambda_c$, is orders of magnitude smaller than the characteristic timescale of the system's macroscopic dynamics, $\tau_s$ \cite{huang2009introduction}. Since $\tau_s \gg \tau_c$, we can take the continuum limit of an infinite collision rate ($\lambda_c \to \infty$) while holding the macroscopic noise intensity constant 
\begin{equation}
\lim_{\lambda_c\to\infty} \lambda_c\langle\theta_i^2\rangle = \text{const}.
\end{equation} 
Correspondingly, individual collisions become infinitesimally weak, with the root-mean-square amplitude scaling as $\sqrt{\langle\theta^2_i\rangle} \propto \sqrt{\tau_c}$. By the Central Limit Theorem \cite{feller1991introduction}, the sum of this immense number of identically distributed random variables converges, transforming the zero-mean white Poisson shot noise into Gaussian white noise $\xi(t)$ \cite{van1983stochastic}, i.e., the formal derivative of the Wiener process \cite{feller1991introduction}. 

Conventionally, the intensity $D_T$ is factored out of the stochastic variable. The effective thermal fluctuations are thus written as $\sqrt{2D_T}\xi(t)$, which redefines $\xi(t)$ as dimensionless Gaussian white noise with zero mean and unit variance:
\begin{equation}
    \langle \xi(t)\rangle = 0,
\end{equation}
\begin{equation}
    \langle\xi(t)\xi(s)\rangle = \delta(t-s).
\end{equation}
Finally, the Fluctuation-Dissipation Theorem bridges these random thermal fluctuations and the systematic dissipation of energy. For a thermal bath at temperature $T$, the noise intensity $D_T$ is directly proportional to the viscous friction $\gamma$ \cite{kubo1966fluctuation}:
\begin{equation}
    D_T = \gamma k_B T,
\end{equation}
where $k_B$ is the Boltzmann constant. Note that $D_T$ here is not the diffusion coefficient. 

The motion of a single overdamped Brownian particle is described by the following Langevin equation:
\begin{equation}
    \gamma \dot{x}(t)=\sqrt{2\gamma k_BT}\xi(t).
    \label{Langevin_simplest}
\end{equation}
The corresponding Fokker-Planck equation governs the time evolution of the probability distribution $P(x,t)$ of the particle's position $x$ \cite{risken1989fokker}:
\begin{equation}
    \frac{\partial P(x,t)}{\partial t}=D\frac{\partial^2P(x,t)}{\partial x^2},
    \label{FP}
\end{equation}
where 
\begin{equation}
D=\frac{k_B T}{\gamma}
\end{equation}
 is the free diffusion coefficient.

Given the probability distribution $P(x(t),t)$ and the trajectory $x(t)$, one can define the stochastic entropy of the particle (or system) \cite{seifert2005entropy,seifert2012stochastic}:
\begin{equation}
    s_{sys}(t)=-k_B \ln P(x(t),t).
    \label{s_sys}
\end{equation}
The particle's entropy, $s_{sys}$, does not represent the total entropy:
\begin{equation}
s_{tot}= s_{sys}+ s_{med},
\end{equation}
where $s_{med}$ denotes the change in the entropy of the medium (or environment) arising from dissipated heat, defined simply as the heat transferred divided by the temperature $T$. Within the trajectory description, the medium entropy at time $t$ (relative to the initial state) is given by \cite{sekimoto2010stochastic}:
\begin{equation}
s_{med}(t)=\frac{Q(t)}{T}=\frac{1}{T}\int _{0}^{t}\left[\gamma \dot{x}(t')-\sqrt{2 \gamma k_B T}\xi (t' )\right]\circ dx(t' ),
\end{equation}
where $Q(t)$ is the total heat transferred from the particle to the environment from the start until time $t$, and the term $(\gamma \dot{x}-\sqrt{2\gamma k_B T}\xi(t))$ represents the force exerted by the particle on the environment. For the system described by Eq.~(\ref{Langevin_simplest}), the stochastic entropy can be readily calculated by solving Eq.~(\ref{FP}). Assuming the particle is initially located at $x=0$ at time $t=0$, the probability distribution at time $t$ is a Gaussian distribution:
\begin{equation}
P(x,t)=\frac{1}{\sqrt{4\pi D t}}\exp \left(-\frac{x^{2}}{4D t}\right).
\end{equation} 
Substituting this expression into Eq.~(\ref{s_sys}) and taking the average yields the average system entropy, $\langle s_{sys}(t)\rangle$, which increases logarithmically over time:
\begin{equation}
\langle s_{sys}(t)\rangle =k_{B}\left[\frac{1}{2}+\ln \sqrt{4\pi D t}\right],
\end{equation}
where we have used the relation $\langle x^2(t)\rangle=2D t$. This logarithmic growth reflects the increasing uncertainty in the particle's position due to diffusion.

Conversely, the entropy of the medium is exactly zero ($s_{med}(t)=0$), because:
\begin{equation}
\gamma \dot{x}=\sqrt{2 \gamma k_B T}\xi(t) \implies \gamma \dot{x}-\sqrt{2 \gamma k_B T}\xi (t)=0.
\end{equation}
At the trajectory level, neither the free particle nor the environment performs macroscopic work on the other; they merely exchange momentum. Consequently, on average, no net heat is exchanged with the reservoir:
\begin{equation}
\langle s_{med}(t)\rangle=0.
\end{equation}
The average total entropy is therefore:
\begin{equation}
 \langle s_{tot}(t)\rangle = \langle s_{sys}(t)\rangle+\langle s_{med}(t)\rangle=k_{B}\left[\frac{1}{2}+\ln \sqrt{4\pi D t}\right].
\end{equation}
Differentiating $\langle s_{tot}(t)\rangle$ with respect to time $t$ yields the average entropy production rate:
\begin{equation}
\langle \dot{s}_{tot}(t)\rangle=\frac{k_B}{2t}.
\end{equation}
Note that the entropy production rate vanishes at infinite time ($\lim_{t\to\infty}\langle \dot{s}_{tot}(t)\rangle=0$), as the probability distribution broadens indefinitely and the system approaches a stationary state. Since $t > 0$, the entropy production rate remains strictly positive ($\langle \dot{s}_{tot}(t)\rangle>0$), in full agreement with the Second Law of Thermodynamics.

In summary, thermal fluctuations satisfy detailed balance and are therefore modeled as zero-mean, symmetric, stationary Gaussian process as a consequence of the Central Limit Theorem. They obey the Fluctuation-Dissipation Theorem and do not produce any medium entropy $\langle \dot{s}_{med}(t) \rangle=0$. In general, thermal fluctuations may not be $\delta$-correlated and can possess memory, in which case the Gaussian noise $\xi(t)$ is no longer white \cite{franosch2011resonances}.

\section{Active Fluctuations}

Not all physical fluctuations have thermal origin; those arising from distinct mechanisms are termed \textit{athermal} fluctuations \cite{kanazawa2017statistical}. They manifest across a wide range of biological \cite{ben2011effective}, electrical \cite{blanter2000shot}, and granular systems \cite{behringer2019physics}. A key subset of these dynamics, specifically those associated with self-propelled systems, is designated as \textit{active fluctuations}. Such fluctuations can originate directly from the stochastic nature of self-propulsion mechanisms, including catalytic reactions in artificial microswimmers \cite{golestanian2005propulsion,howse2007self} or ATP hydrolysis in molecular motors like dynein, kinesin, and myosin \cite{gilbert1995pathway,ross2006processive,prost2015active}. Another prominent source is the so-called \textit{active bath}, composed of active matter suspensions \cite{seyforth2022nonequilibrium,ye2020active}, where frequent collisions with self-propelled agents yield dynamics that differ fundamentally from those of a classical thermal bath. Finally, active fluctuations can also emerge as indirect consequences of self-propulsion, driven either by localized heat production from chemical reactions \cite{mabillard2023heat,hwang2017quantifying} or by hydrodynamic flows generated as active matter navigating a viscoelastic medium \cite{plan2020active}.

Let us retrace the core arguments from the previous chapter and compare the characteristics of thermal and active fluctuations. We denote the stochastic process modeling active fluctuations as $\eta_A(t)$. Consider a particle subjected solely to this active force:
\begin{equation}
\gamma\dot{x}(t)=\eta_A(t).
\label{Langevin_simplest_act}
\end{equation}

Unlike thermal noise, active fluctuations are inherently out of equilibrium and break detailed balance \cite{gnesotto2018broken}. This microscopic irreversibility often stems from asymmetric fluctuation statistics: active forces may exhibit a non-zero mean 
\begin{equation}
\langle\eta_A(t)\rangle \neq 0,
\end{equation}
 possess a heavy tail in a preferred direction, or feature a zero mean $\langle\eta_A(t)\rangle = 0$ alongside a non-zero skewness. It is worth noting, however, that while active fluctuations are frequently non-Gaussian, certain theoretical models of $\eta_A(t)$ retain Gaussian statistics.

Furthermore, active fluctuations are completely decoupled from the system's viscous friction $\gamma$. Because they are sustained by external or internal energy sources, they violate the Fluctuation-Dissipation Theorem \cite{dabelow2019irreversibility,marconi2008fluctuation,maggi2017memory}. The intensity of active noise can depend on parameters such as fuel concentration \cite{needleman2017active}, the efficiency of the underlying self-propulsion mechanism, or the local density of active agents. Active fluctuations may also possess memory, characterized by a non-delta autocorrelation function:
\begin{equation}
\langle \eta_A(t)\eta_A(s)\rangle = C R(t-s),
\end{equation}
where $R(t-s) \neq \delta(t-s)$ and $C$ is some constant.

Continuous energy consumption by self-propulsion drives continuous entropy production. In certain systems, this entropy production rate can be calculated explicitly \cite{dabelow2019irreversibility,huang2026entropy}. Similarly to thermal systems, the average system entropy increases over time due to spatial diffusion, with its rate of change vanishing in the long-time limit, $\langle \dot{s}_{\text{sys}}(t)\rangle \propto 1/t$. Conversely, the medium entropy $\langle s_{\text{med}}(t)\rangle$ steadily increases because active fluctuations generate heat via the force exerted by the particle on the fluid:
\begin{equation}
dQ = \gamma \dot{x} \circ dx = \gamma \left(\frac{\eta_A(t)}{\gamma}\right)\left(\frac{\eta_A(t)}{\gamma}dt\right) = \frac{\eta_A^2(t)}{\gamma}dt.
\end{equation}
The rate of medium entropy production is therefore given by:
\begin{equation}
\dot{s}_{\text{med}}(t) = \frac{1}{T} \frac{dQ}{dt} = \frac{\eta^2_A(t)}{\gamma T}.
\end{equation}
Because $\eta_A^2(t) \ge 0$, the medium entropy strictly increases. In the long-time limit ($t \to \infty$), the average total entropy production rate approaches a constant non-zero value:
\begin{equation}
\lim_{t\to\infty}\langle \dot{s}_{\text{tot}}(t)\rangle = \frac{\langle \eta^2_A(t)\rangle}{\gamma T}.
\end{equation}
Note that this derivation applies directly to the simplified dynamics in Eq.~\eqref{Langevin_simplest_act}.

The average total entropy production rate quantifies the thermodynamic irreversibility of active systems. While a vanishing rate ($\langle \dot{s}_{\text{tot}}\rangle = 0$) indicates time-reversibility \cite{mandal2017entropy}, generic active systems exhibit positive entropy production and break detailed balance \cite{dabelow2019irreversibility,o2022time,gnesotto2018broken}.

The key operational differences between thermal and active fluctuations are summarized in Table~\ref{table_one}. Below, we present the principal stochastic processes used to model active fluctuations $\eta_A(t)$.

\begin{table}[t]
\centering
\renewcommand{\arraystretch}{1.5}
\begin{tabular}{|C{0.3\textwidth}|C{0.3\textwidth}|C{0.3\textwidth}|}
\hline
\textbf{Property / Quantity} & \textbf{Thermal Noise $\xi(t)$} & \textbf{Active Noise $\eta_A(t)$} \\
\hline
Detailed balance & Satisfied & Violated \\
\hline
Symmetry & Symmetric,\par $\rho(\theta) = \rho(-\theta)$ & Arbitrary\par (can be asymmetric) \\
\hline
Probability Density & Gaussian distribution & Non-Gaussian in general \\ 
\hline
Mean Value & Zero mean\par $(\langle \xi(t) \rangle = 0)$ & Arbitrary\par $(\langle \eta_A(t) \rangle \neq 0\text{ in general})$ \\ 
\hline
Fluctuation-Dissipation\par Theorem & Satisfied & Violated \\
\hline
Asymptotic Entropy\par Production Rate\par $\lim_{t\to\infty}\langle\dot{s}_{\text{tot}}\rangle$ & Vanishes & Approaches positive constant \\ 
\hline
\end{tabular}
\caption{Comparison of thermal and active fluctuations.}
\label{table_one}
\end{table}

\subsection{Active Ornstein-Uhlenbeck Process}
\label{subsec:aoup}

The Ornstein-Uhlenbeck process, originally developed to model velocity fluctuations in inertial Brownian motion \cite{uhlenbeck1930theory}, has become a foundational model in active matter research \cite{dabelow2019irreversibility,mandal2017entropy,martin2021statistical}. An active Ornstein-Uhlenbeck particle models either an active agent driven by a persistent self-propulsion force or a passive particle subjected to collisions in a dense active bath \cite{martin2021statistical,nguyen2022active,woillez2020nonlocal,nagai2015collective}. An Ornstein-Uhlenbeck process $\eta_{\text{OU}}(t)$ with mean $\mu_{\text{OU}}$ is governed by the Langevin equation \cite{fodor2016far}:
\begin{equation}
    \tau \dot{\eta}_{\text{OU}}(t) = -(\eta_{\text{OU}}(t) - \mu_{\text{OU}}) + \sqrt{2D_{\text{OU}}} \zeta(t),
    \label{ou_langevin}
\end{equation}
where $\zeta(t)$ is a zero-mean, delta-correlated Gaussian white noise process. Here, $\tau$ denotes the persistence time, and $D_{\text{OU}}$ represents the effective noise intensity. Equation~\eqref{ou_langevin} admits the exact analytical solution:
\begin{equation}
    \eta_{\text{OU}}(t) = \mu_{\text{OU}} + (\eta_0 - \mu_{\text{OU}})e^{-(t-t_0)/\tau} + \frac{\sqrt{2D_{\text{OU}}}}{\tau} \int_{t_0}^t e^{-(t-s)/\tau} \zeta(s) \, ds,
\end{equation}
where $\eta_{\text{OU}}(t_0) = \eta_0$ is the initial condition at $t_0$. In the stationary regime ($t \gg \tau$), the mean and autocovariance are \cite{fodor2016far}:
\begin{align}
    \langle \eta_{\text{OU}}(t) \rangle &= \mu_{\text{OU}}, \\
    \langle [\eta_{\text{OU}}(t) - \mu_{\text{OU}}][\eta_{\text{OU}}(s) - \mu_{\text{OU}}] \rangle &= \frac{D_{\text{OU}}}{\tau} \exp\left( -\frac{|t - s|}{\tau} \right).
\end{align}
The finite persistence time $\tau$ fundamentally distinguishes the Onstein-Uhlenbeck noise $\eta_{\text{OU}}(t)$ from white Gaussian noise $\xi(t)$ by rendering the system non-Markovian; that is, future dynamics depend not only on the instantaneous state but on past trajectory history. In the limit of vanishing persistence time ($\tau \to 0$), zero-mean Ornstein-Uhlenbeck noise converges to Gaussian white noise \cite{haunggi1994colored}:
\begin{equation}
    \lim_{\tau\to0}\eta_{\text{OU}}(t)=\sqrt{2D_{\text{OU}}}\zeta(t).
\end{equation}

\begin{figure}[t]
    \centering
    \includegraphics[width=0.49\textwidth]{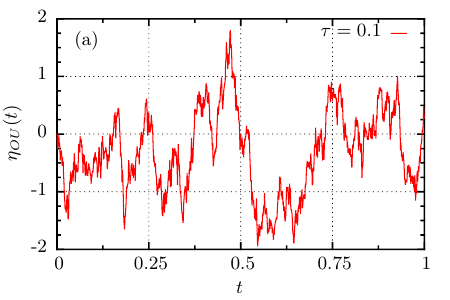}
    \includegraphics[width=0.49\textwidth]{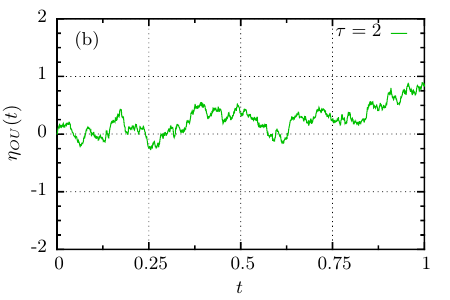} 
    \caption{Representative realizations of a zero-mean ($\mu_{\text{OU}} = 0$) Ornstein-Uhlenbeck process $\eta_{\text{OU}}(t)$ for two correlation times: $\tau=0.1$ (a) and $\tau=2.0$ (b).}
    \label{fig:ornstein}
\end{figure}

\subsection{White Poisson Shot Noise}
\label{subsec:shot_noise}

Certain active mechanisms, such as discrete ATP hydrolysis events, can be modeled as instantaneous energy inputs \cite{mabillard2023heat,fiasconaro2009tuning,shen2005nonequilibrium}. When such events are sparse, the active force $\eta_A(t)$ vanishes between successive firing events. To describe discrete, $\delta$-like impulses occurring at random Poissonian times, we employ white Poisson shot noise, defined analogously to microscopic thermal collisions in Eq.~\eqref{poisson_thermal} \cite{hanggi1978derivations,hanggi1980langevin}:
\begin{equation}
    \eta_P(t) = \sum_{i=1}^{n(t)} z_i \delta(t - t_i).
\end{equation}
Here, $n(t)$ is a Poisson counting process \cite{feller1991introduction} with a mean spiking rate $\lambda > 0$ determining the arrival times $t_i$. The jump amplitudes $\{z_i\}$ are independent and identically distributed random variables drawn from a probability density function $\rho(z)$. Crucially, Poisson shot noise differs from microscopic thermal fluctuations in that the average interval between pulses, $\tau_P = 1/\lambda$, need not be small, nor must the amplitude distribution $\rho(z)$ be Gaussian. Its covariance is
\begin{equation}
 \langle \eta_P(t) \eta_P(s)\rangle -\langle \eta_P(t)\rangle \langle\eta_P(s)\rangle = 2D_P\delta(t-s),
\end{equation}
where $D_P=\lambda \langle z_i^2\rangle/2$ is Poisson noise intensity.
 For overdamped motion driven by Eq.~\eqref{Langevin_simplest_act}, the corresponding probability density $P(x,t)$ obeys the integro-differential Fokker-Planck-Kolmogorov-Feller equation \cite{hanggi1978derivations,hanggi1980langevin}:
\begin{equation}
    \frac{\partial P(x,t)}{\partial t} = \lambda \int_{-\infty}^{\infty} \left[ P\left(x-\frac{z}{\gamma},t\right) - P(x,t) \right] \rho(z) \, dz.
    \label{FPKF}
\end{equation}

\begin{figure}[t]
    \centering
    \includegraphics[width=0.49\textwidth]{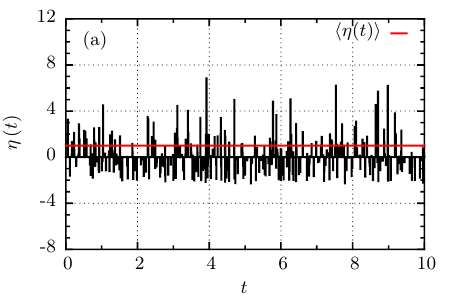}
    \includegraphics[width=0.49\textwidth]{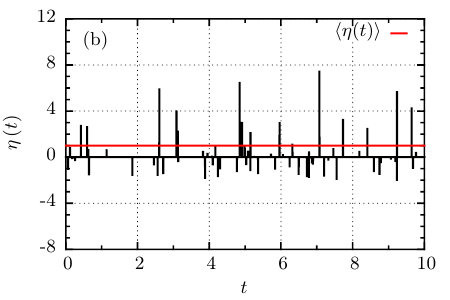}
    \caption{Representative trajectories of white Poisson shot noise with a skew-normal amplitude distribution $\rho_s(z)$ for different mean spiking rates $\lambda$ and average amplitudes $\zeta$. The solid red line indicates the mean bias $\langle\eta_P(t)\rangle=1$. Plot reproduced from \hyperref[A1]{A1}.}
    \label{fig:poisson}
\end{figure}

\begin{figure}[t]
    \centering
    \includegraphics[width=0.49\textwidth]{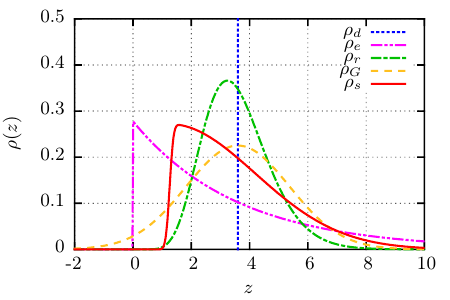} 
    \caption{Comparison of probability density functions $\rho(z)$ for active Poissonian shot noise amplitudes, ranging from unidirectional models to bidirectional skew-normal statistics. Plot reproduced from \hyperref[A2]{A2}.}
    \label{fig:distributions}
\end{figure}

The choice of $\rho(z)$ reflects the specific underlying active driving process. Figure~\ref{fig:distributions} illustrates several representative amplitude distributions. These distributions are characterized by their lowest central statistical moments: mean $\zeta$, variance $\sigma^2$, and skewness $\chi$ \cite{feller1991introduction}:
\begin{equation}
    \zeta = \langle z_i \rangle,
\end{equation}
\begin{equation}
    \sigma^2 = \langle (z_i - \zeta)^2 \rangle,
\end{equation}
\begin{equation}
    \chi = \frac{\langle (z_i - \zeta)^3 \rangle}{\sigma^3}.
\end{equation}

The simplest configuration assumes constant impulse magnitudes, serving as a deterministic reference limit where amplitude randomness vanishes:
\begin{equation}
    \rho_d(z) = \delta(z-\zeta).
\end{equation}

A standard fully stochastic continuous model employs the exponential distribution \cite{chowdhury2013stochastic}:
\begin{equation}
    \rho_{e}(z) = \frac{\theta(z)}{\zeta} \exp\left(-\frac{z}{\zeta}\right),
    \label{pdf_exp}
\end{equation}
where $\theta(z)$ is the Heaviside step function. This distribution is strictly monotonic, with variance tied directly to its mean ($\sigma^2_e = \zeta^2$) and a fixed positive skewness ($\chi = 2$).

When a specific nonzero impulse amplitude is preferred, a non-monotonic distribution is required. A natural generalization of the exponential distribution is the Erlang distribution \cite{johnson1994}:
\begin{equation}
    \rho_{r}(z,n) = \frac{\theta(z) z^{n-1} n^n}{\zeta^n (n-1)!} \exp\left(-\frac{n z}{\zeta}\right).
    \label{pdf_erlang}
\end{equation}
The sum of $n$ independent exponentially distributed variables with mean $\mu = \zeta/n$ follows an Erlang distribution. For $n > 1$, $\rho_r(z,n)$ is non-monotonic, attaining a peak at $z = (n-1)\zeta/n$. Higher values of $n$ lead to faster-decaying tails and reduced asymmetry, with skewness given by $\chi = 2/\sqrt{n}$.

The preceding models enforce purely positive impulses ($z_i \ge 0$). However, real active fluctuations can exert bidirectional forces. To capture bidirectional fluctuations while retaining control over variance and mean, we consider the Gaussian amplitude distribution \cite{johnson1994}:
\begin{equation}
    \rho_G(z,\sigma_G^2) = \frac{1}{\sqrt{2\pi \sigma_G^2}} \exp\left( -\frac{(z-\zeta)^2}{2\sigma^2_G} \right).
\end{equation}
While individual amplitude statistics are normal (as in thermal noise), the resulting shot-noise process $\eta_P(t)$ possesses a non-zero mean force and a net intensity $D_P = \lambda \langle z_i^2 \rangle / 2$ governed by the rate $\lambda$. The Gaussian distribution is symmetric about $\zeta$, yielding zero skewness ($\chi = 0$).

When active systems feature multiple concurrent driving mechanisms, symmetric amplitude assumptions ($\rho(\zeta+\epsilon) = \rho(\zeta-\epsilon)$) are often insufficient. To incorporate controlled asymmetry into bidirectional fluctuations, we utilize the skew-normal distribution \cite{azz,rijal2022exact}:
\begin{equation}
    \rho_s(z) = \frac{2}{\sqrt{2\pi \omega^2}} \exp\left(-\frac{(z-\mu)^2}{2\omega^2}\right) \int_{-\infty}^{\alpha[(z-\mu)/\omega]} \frac{1}{\sqrt{2\pi}} \exp\left(-\frac{s^2}{2}\right) ds,
\end{equation}
where $\mu$, $\omega$, and $\alpha$ are location, scale, and shape parameters. These are uniquely specified by the target mean $\zeta$, variance $\sigma^2$, and skewness $\chi$ \cite{generacja,generacja2}:
\begin{equation}
\alpha = \frac{\delta}{\sqrt{1-\delta^2}}, 
\label{alpha}
\end{equation}
\begin{equation}
\omega = \sqrt{\frac{\sigma^2}{1- 2\delta^2/\pi}},
\end{equation}
\begin{equation}
\mu = \zeta - \delta\sqrt{\frac{2\sigma^2}{\pi(1-2\delta^2/\pi)}},
\label{eq_S_def}
\end{equation}
where $\delta$ is defined as:
\begin{equation}
    \delta = \text{sgn}(\chi) \sqrt{\frac{|\chi|^{2/3}}{(2/\pi)\left\{[(4-\pi)/2]^{2/3} + |\chi|^{2/3}\right\}}}.
    \label{eq_S_delta}
\end{equation}
Because $|\delta| < 1$ in Eq.~\eqref{alpha}, the skewness parameter is bounded within the range:
\begin{equation}
    |\chi| < \left[ \frac{2\left(\frac{4-\pi}{2}\right)^{2/3}}{\pi-2} \right]^{3/2} \approx 0.995.
\end{equation}
This distribution constitutes a cornerstone of the subsequent chapters of this thesis.

\subsection{Other Models of Active Fluctuations}
\label{subsec:other_fluct}

Several other stochastic processes are used to model active driving beyond the active Ornstein-Uhlenbeck process and white Poisson shot noise. Prominent among them is colored Poisson noise, which replaces instantaneous $\delta$-impulses with pulses of finite temporal duration \cite{paneru2023bona,lee2022effects}. It is generated in a similar way to Ornstein-Uhlenbeck process by an overdamped Langevin equation with Poisson shot noise $\eta_P(t)$ replacing the Gaussian white noise $\xi(t)$ in Eq.~(\ref{ou_langevin}):
\begin{equation}
\tau \dot{\eta}_{CP}(t) = -(\eta_{CP}(t) - \langle \eta_{CP}(t)\rangle ) + \eta_P(t).
\end{equation}
The mean and covariance of the colored Poisson noise follow
\begin{equation}
\langle \eta_{CP}(t)\rangle=\lambda \tau \langle z_i\rangle,
\end{equation}
\begin{equation}
\langle \eta_{CP}(t) \eta_{CP}(s)\rangle -\langle \eta_{CP}(t)\rangle \langle\eta_{CP}(s)\rangle = \frac{\lambda \tau \langle z_i^2\rangle}{2}\exp\left(-\frac{|t-s|}{\tau}\right),
\end{equation}
which means that colored Poisson noise is exponentially correlated.

Another useful process is dichotomous noise (or telegraphic noise), which models discrete stochastic switching between active states, such as run-and-tumble motion or reorientation events \cite{romanczuk2012active,mondal2025role}.
Asymmetric dichotomous noise, $\eta_D(t)$, is a two-state Markovian jump process alternating between active force levels $\eta_+$ and $\eta_-$. Transitions are governed by independent Poisson rates $\nu_+$ (for $\eta_+ \to \eta_-$) and $\nu_-$ (for $\eta_- \to \eta_+$). The stationary probabilities $P_\pm$ and mean active force $\eta_D$ are:
\begin{equation}
    P_+ = \frac{\nu_-}{\nu_+ + \nu_-}, \quad P_- = \frac{\nu_+}{\nu_+ + \nu_-}
\end{equation}
\begin{equation}
 \langle \eta_D(t)\rangle = \frac{\eta_+ \nu_- + \eta_- \nu_+}{\nu_+ + \nu_-}.
\end{equation}
Dichotomous noise is also exponentially correlated, its covariance reads
\begin{equation}
\langle \eta_D(t) \eta_D(s)\rangle -\langle \eta_D(t)\rangle \langle\eta_D(s)\rangle = \frac{\nu_+\nu_-(\eta_+-\eta_-)^2}{(\nu_++\nu_-)^2}\exp(-(\nu_++\nu_-)|t-s|).
\end{equation}
 consequently, systems driven by colored Poisson and dichotomous noise are non-Markovian. Under certain limits, these processes reduce to white Poisson shot noise or Gaussian white noise.

\begin{figure}[t]
    \centering
    \includegraphics[width=0.98\textwidth]{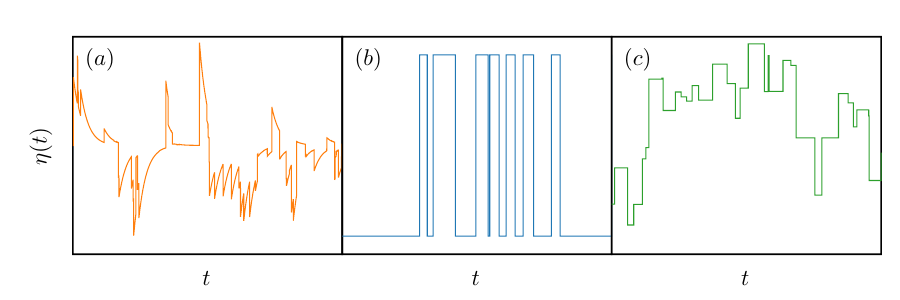} 
    \caption{Exemplary realizations of the colored Poisson noise $(a)$, dichotomous noise $(b)$ and Kubo-Anderson fluctuations $(c)$.}
    \label{fig:dck}
\end{figure}

A final noteworthy model is the Kubo-Anderson process $\eta_{KA}(t)$, a specialized subclass of the purely discontinuous Kolmogorov-Feller kangaroo process and a generalization of the dichotomous noise with a constant jump frequency $\lambda$ \cite{vilela2024dynamics,luczka2005non}. The transition probability rate $\mathcal{W}(z|z_0)$ from state $z_0$ to $z$ factorizes as:
\begin{equation}
    \mathcal{W}(z|z_0) = \lambda Q(z),
\end{equation}
where $Q(z)$ is the normalized probability of jumping to state $z$. In a general kangaroo process, the jump rate depends on the initial state, $\lambda = \lambda(z_0)$. The Kubo-Anderson process displays exponential autocorrelation dynamics. Its average and second moment are determined by the probability distribution $Q(z)$
\begin{equation}
\langle \eta_{KA}(t)\rangle=\int_{-\infty}^{\infty} z Q(z)dz,
\end{equation}
\begin{equation}
\langle \eta^2_{KA}(t)\rangle=\int_{-\infty}^{\infty} z^2 Q(z)dz.
\end{equation}
Then the covariance reads
\begin{equation}
\langle \eta_{KA}(t) \eta_{KA}(s)\rangle -\langle \eta_{KA}(t)\rangle \langle\eta_{KA}(s)\rangle = (\langle \eta^2_{KA}(t)\rangle-\langle \eta_{KA}(t)\rangle^2)\exp(-\lambda|t-s|).
\end{equation}

\section{Transport Phenomena}
Before investigating the dynamics of a system driven by active fluctuations, a benchmark for comparison is required. The simplest approach to driving a system out of equilibrium is by applying a constant external force $F$. The dynamics of a free Brownian particle of mass $m$ is described by the Langevin equation:
\begin{equation}
    m\ddot{x}(t)+\gamma \dot{x}(t)=F+\sqrt{2\gamma k_BT}\xi(t).
\label{potential_bez}
\end{equation}
In the absence of bias ($F=0$), the system relaxes to a Gibbs equilibrium state \cite{gardiner1985handbook}. If $F \neq 0$, the system reaches a nonequilibrium steady state (NESS) in the long-time limit ($t\to\infty$) \cite{seifert2012stochastic}, wherein the particle moves with a constant long-time average velocity $\langle v\rangle$. We define this quantity as the primary quantifier of transport throughout this thesis:
\begin{equation}
    \langle v\rangle = \lim_{t\to\infty}\frac{\langle x(t)-x(0)\rangle}{t}.
    \label{av_v}
\end{equation}
For the free particle described by Eq.~\eqref{potential_bez}, this velocity depends solely on the magnitude of the force $F$ and the friction parameter $\gamma$:
\begin{equation}
    \langle v\rangle=\frac{F}{\gamma}=\mu F=v_0,
    \label{free_v}
\end{equation}
where $\mu$ denotes the particle's mobility, defined as:
\begin{equation}
    \mu(F)=\frac{\langle v\rangle(F)}{F}.
\end{equation}
In the weak-force limit, this quantity defines the \textit{absolute mobility}:
\begin{equation}
    \mu_0=\lim_{F\to0}\frac{\langle v\rangle(F)}{F}.
\end{equation}
When an equilibrium system is subjected to a vanishingly small perturbation, the average velocity $\langle v\rangle$ scales linearly with force, placing the system within the linear response regime.

In this work, we focus on particle dynamics constrained by an additional periodic potential $U(x) = U(x+L)$ of spatial period $L$. In this setting, the Langevin equation~\eqref{potential_bez} takes the form \cite{risken1989fokker,reimann2002brownian,hanggi2009artificial}:
\begin{equation}
        m\ddot{x}(t)+\gamma\dot{x}(t)=-U'(x)+F+\sqrt{2\gamma k_BT}\xi(t),
\label{const_pot}
\end{equation}
where the prime denotes spatial differentiation with respect to $x$. Brownian motion in a periodic potential serves as a foundational paradigm in statistical mechanics with remarkable universality \cite{spiechowicz2016transient,spiechowicz2022diffusion}. Equation~\eqref{const_pot} is mathematically isomorphic to a wide array of physical systems, including the noisy pendulum \cite{gitterman2008noisy}, superionic conductors \cite{risken1989fokker}, Josephson junctions \cite{fulde1975problem}, phase-locked loops \cite{viterbi1963phase}, rotating dimers \cite{bellando2022giant}, fluxon dynamics in superconductors \cite{falo1999ratchet}, charge density waves \cite{gruner1981nonlinear}, and the Frenkel-Kontorova model \cite{braun1998nonlinear}.

The parameter space of Eq.~\eqref{const_pot} is inherently high-dimensional. Casting the equation into dimensionless form reduces the number of independent parameters to a minimum and renders theoretical predictions independent of specific experimental realizations. This universality is particularly advantageous when verifying theoretical predictions experimentally, as one can choose an accessible experimental platform to bypass technical constraints present in other setups. We introduce the dimensionless position $\hat{x}$ and time $\hat{t}$ as:
\begin{equation}
    \hat{x}=\frac{2\pi x}{L}, \quad \hat{t}=\frac{t}{\tau_0},
\end{equation}
\begin{equation}
    \tau_0=\frac{\gamma L^2}{4\pi^2 k_BT},
\end{equation}
where $\tau_0$ represents the characteristic diffusive timescale across a spatial period $L/(2\pi)$. Applying the chain rule, the derivatives transform as follows:
\begin{equation}
    \ddot{x}(t)=\frac{L}{2\pi\tau_0^2}\ddot{\hat{x}}(\hat{t}),
\end{equation}
\begin{equation}
    \dot{x}(t)=\frac{L}{2\pi\tau_0}\dot{\hat{x}}(\hat{t}),
\end{equation}
\begin{equation}
    U'(x)=\frac{2\pi}{L}\hat{U}'(\hat{x}).
\end{equation}

Substituting these relations into Eq.~\eqref{const_pot} yields the dimensionless Langevin equation:
\begin{equation}
    \hat{m}\ddot{\hat{x}}(\hat{t})+\dot{\hat{x}}(\hat{t})=-\hat{U}'(\hat{x})+\hat{F}+\sqrt{{2D_T}}\hat{\xi}(\hat{t}),
    \label{const_dimensionless}
\end{equation}
where the dimensionless mass $\hat{m}$ is defined as:
\begin{equation}
    \hat{m}=\frac{m}{\gamma\tau_0}=\frac{4\pi^2mk_BT}{\gamma^2L^2},
\end{equation}
the dimensionless friction coefficient is set to unity ($\hat{\gamma}=1$), the dimensionless potential $\hat{U}(\hat{x})$ is:
\begin{equation}
    \hat{U}(\hat{x})=\frac{1}{k_B T}U\left(\frac{L\hat{x}}{2\pi}\right),
\end{equation}
the dimensionless constant force $\hat{F}$ is:
\begin{equation}
    \hat{F}=\frac{L}{2\pi k_BT}F,
\end{equation}
and $\hat{\xi}(\hat{t})$ represents dimensionless Gaussian white noise:
\begin{equation}
    \hat{\xi}(\hat{t})=\frac{L}{2\pi k_B T}{\xi}(\tau_0\hat{t}).
\end{equation}
Under this rescaling, the effective thermal noise intensity equals unity:
\begin{equation}
    D_T=\frac{4\pi^2k_BT\tau_0}{\gamma L^2}=1.
\end{equation}

For brevity, we omit the hat notation in subsequent chapters. This scaling is suited for analyzing the strong-damping regime, where viscous friction dominates inertial effects. In the limit $m \to 0$, the inertial term $m\ddot{x}(t)$ vanishes, reducing Eq.~\eqref{const_dimensionless} to the overdamped first-order Langevin equation:
\begin{equation}
    \dot{x}(t)=-U'(x)+F+\sqrt{2D_T}\xi(t).
    \label{overdamped_dimensionless}
\end{equation}
Overdamped dynamics underpin the study of intracellular transport along microtubules \cite{reimann2002brownian,bier2003processive,reimann2002introduction} and the motion of microswimmers \cite{ten2011brownian,volpe2014simulation}. In this limit, Stratonovich derived the exact analytical expression for the stationary average velocity \cite{risken1989fokker,stratonovich1967topics}:
\begin{equation}
    \langle v\rangle = \frac{L D_T \left(1 - e^{-L F / D_T}\right)}{\int_0^{L} dx \int_0^{L} dz \, \exp\left[ \frac{U(x+z) - U(x) - Fz}{D_T} \right]}.
    \label{stratonov}
\end{equation}

\begin{figure}[t]
    \centering
    \includegraphics[width=0.49\linewidth]{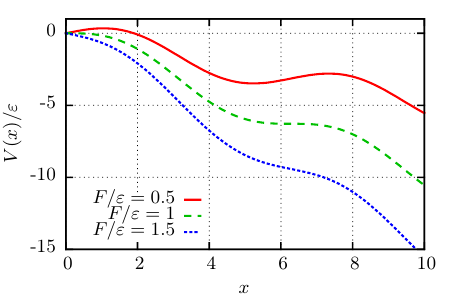}
    \includegraphics[width=0.49\linewidth]{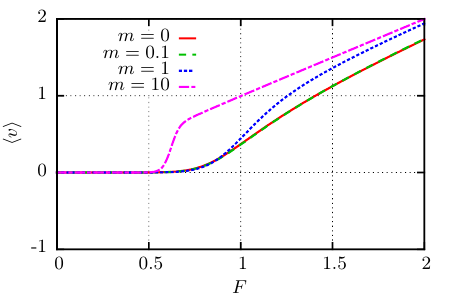}
    \caption{Generalized tilted washboard potential $V(x)$ (left panel). Average velocity $\langle v\rangle$ as a function of external bias $F$ for different mass values $m$, with thermal noise intensity $D_T=0.1$ and barrier parameter $\varepsilon=1$ (right panel).}
    \label{fig:deterministic}
\end{figure}
Novel analytical methods for this system are continually emerging. Recently, the intermediate scattering function was used to calculate the values of quantities of interest   \cite{rusch2026intermediate}. If the periodic potential vanishes, $U(x)=0$, the average velocity of a free particle is recovered, $\langle v\rangle=F$. The same holds true when the potential barrier is much smaller than the driving force, $\Delta U\ll F$, causing the potential minima to disappear; under these conditions, nothing hinders the Brownian particle, and its average velocity approaches that of a free Brownian particle, $\langle v\rangle\to v_0$. To better illustrate this behavior, let us consider a simple, specific form of the potential, $U(x)=\varepsilon\sin(x)$, which has a period of $L=2\pi$ and a barrier height of $\Delta U=2\varepsilon$. The static bias $F$ is usually integrated into $U(x)$ to form the generalized potential, which is often called a tilted washboard potential
\begin{equation}
    V(x)=U(x)-Fx.
\end{equation}
The Langevin equation then takes the form:
\begin{equation}
    \dot{x}(t)=-V'(x)+\sqrt{2D_T}\xi(t).
    \label{overdamped_general}
\end{equation}

To better understand the motion of the particle described by Eq.~(\ref{overdamped_general}), see Fig.~\ref{fig:deterministic}, where $V(x)$ is plotted. Let us first consider a fully deterministic system (i.e., in the absence of thermal fluctuations $\xi(t)$). If the external force is smaller than the potential barrier, $F<\varepsilon$, the generalized potential $V(x)$ exhibits local minima. Because there is no momentum in overdamped dynamics, the particle cannot escape a potential minimum, yielding $\dot{x}(t)=-V'(x)=0$. This state is called the \textit{locked} state. On the other hand, if the bias is larger, $F>\varepsilon$, the potential minima disappear and the particle moves in the direction of the force $F$ with a velocity $\sqrt{F^2-\varepsilon^2}/\gamma$  \cite{mccumber1968effect}. In this case, the particle is in the \textit{running} state.

Inertial dynamics is more complex, particularly in the parameter regime where the potential $V(x)$ possesses local minima ($F<\varepsilon$) and the inertia of the Brownian particle is large enough so that it has sufficient momentum to overcome the potential barrier after passing through a minimum. Then, instead of a simple locked state, both running and locked states can coexist, rendering the system \textit{bistable}  \cite{gruner1981nonlinear} or even multistable \cite{spiechowicz2021arcsine,spiechowicz2022velocity}. Depending on the initial position and velocity, either the locked or the running state is selected. The addition of thermal noise $\xi(t)$ to the system enables switching between these states \cite{spiechowicz2021conundrum}. 
 
To compare the overdamped and inertial dynamics, the dependence of the average velocity $\langle v\rangle$ on the bias $F$ is plotted for different values of the mass $m$ in the right panel of Fig.~\ref{fig:deterministic}. Results in the strong-damping regime, i.e., for negligible inertia ($m=0.1$), are virtually indistinguishable from those of the overdamped system ($m=0$). On the other hand, when the inertial term $m\ddot{x}(t)$ dominates the dynamics ($m\gg1$), the results differ significantly \cite{marchenko2025approach}. The primary distinction is that, for a smaller mass $m$, directed transport ($\langle v\rangle>0$) emerges at a larger bias $F$, and for a given bias $F$, the value of the average velocity $\langle v\rangle$ is smaller for a system with larger relative damping (smaller mass $m$). In addition, in the large-bias regime ($F\gg\varepsilon$), the average velocity $\langle v\rangle$ approaches the value of a free particle, $\langle v\rangle\approx F=v_0$. This limiting value is reached at a smaller bias $F$ when inertia is large.
%

\section{Taxonomy of Transport Anomalies}
\label{sec:Taxonomy}

Transport phenomena that run counter to physical intuition are of significant interest, as uncovering their underlying mechanisms addresses fundamental questions in nonequilibrium dynamics. Consequently, anomalous transport constitutes a central theme of this thesis. Our primary quantity of interest is the stationary average velocity $\langle v\rangle$ defined in Eq.~\eqref{av_v}, which we benchmark against that of a free Brownian particle [Eq.~\eqref{potential_bez}] whose terminal velocity is $v_0 = F/\gamma$ (or $v_0 = F$ in dimensionless units, as per Eq.~\eqref{free_v}).

\begin{figure}[t]
    \centering
    \includegraphics[width=\linewidth]{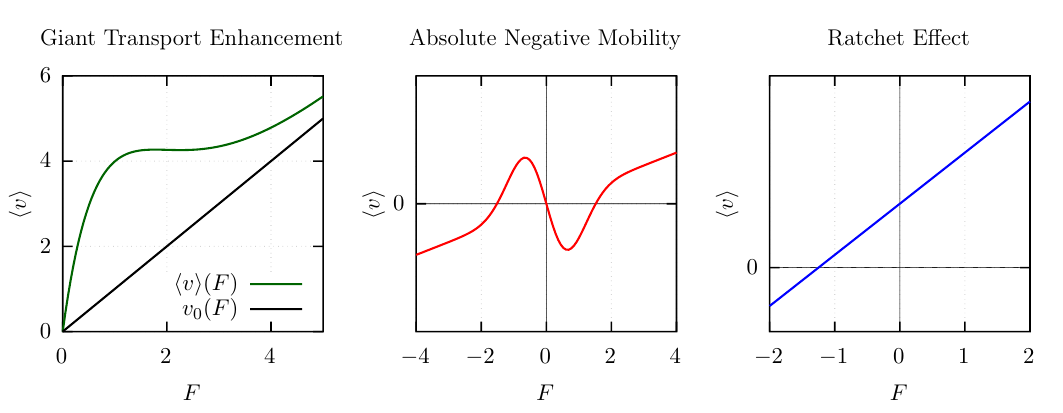}
    \caption{Schematic velocity-force response $\langle v\rangle(F)$ displaying various transport anomalies in comparison to free Brownian motion $v_0=F$.}
    \label{fig:anomalies}
\end{figure}

Anomalous behavior can manifest either in the magnitude $|\langle v\rangle|$ or the direction of transport. In one dimension, direction is uniquely specified by the sign of the average velocity, $\operatorname{sgn}(\langle v\rangle)$. Figure~\ref{fig:anomalies} illustrates representative force-velocity characteristics $\langle v\rangle(F)$ for the transport anomalies considered in this work.

\subsection{Giant Transport Enhancement}

Intuitively, subjecting a Brownian particle to a periodic potential landscape hinders motion due to energy barriers, constraining the average velocity below that of a free particle ($|\langle v\rangle| \le v_0$). In the overdamped regime, this upper bound holds rigorously as a consequence of the Stratonovich formula [Eq.~\eqref{stratonov}] \cite{risken1989fokker,stratonovich1967topics}. Systems wherein the average velocity exceeds the free-particle limit,
\begin{equation}
    \langle v\rangle > v_0 \quad \text{or even} \quad \langle v\rangle \gg v_0,
\end{equation}
are termed \textit{giant transport enhancement}.

Crucially, this enhancement does not stem from additional static forces with non-zero mean, which would merely shift the baseline free-particle velocity $v_0$. Instead, giant transport enhancement typically arises in nonequilibrium conditions induced by time-periodic potential oscillations, such as $U(x,t)=\varepsilon\cos(x+x_0\cos(\omega t))$ or $U(x,t)=\varepsilon (1+\cos(\omega t))\cos(x)$ \cite{mu2009enhanced,romanczuk2010quasideterministic}. External driving continuously injects energy into the system, breaking detailed balance. The dramatic speedup relies on timescale synchronization, for instance, between the potential driving frequency $\omega$ and the intra-well relaxation time. This effect has been experimentally confirmed using optical lattice systems \cite{mu2009enhanced}.

\subsection{Absolute Negative Mobility}

By Newton's second law, a net external force applied to a free particle induces acceleration along the force vector. For a Brownian particle described by Eq.~\eqref{potential_bez} starting from rest ($v(0)=0$), an applied force causes it to accelerate until it reaches terminal velocity settling into a nonequilibrium steady state.

In most physical systems, the absolute mobility $\mu_0$ is strictly positive. When the response opposes the bias,
\begin{equation}
    \mu_0 = \lim_{F\to0}\frac{\langle v\rangle(F)}{F} < 0,
\end{equation}
the system exhibits \textit{absolute negative mobility} (ANM) \cite{ros2005absolute,eichhorn2005moving}. First observed in electrical devices as negative resistance \cite{keay1995dynamic}, ANM has since been demonstrated theoretically and experimentally in Brownian motors \cite{machura2007absolute,speer2007transient,spiechowicz2019coexistence}, colloids in laminar flows \cite{sarracino2016nonlinear}, Josephson junctions \cite{nagel2008observation}, particles moving through corrugated channels \cite{ghosh2014giant}, and active matter \cite{spiechowicz2013absolute,rizkallah2023absolute}. Macroscopic analogues include Brazil nut segregation in vibrated granular media \cite{duran1993arching}. Because ANM is acutely sensitive to particle characteristics, it provides a robust mechanism for species separation by mass or size under identical external driving \cite{slapik2019tunable,slapik2019temperature,slapik2020tunable}.

Historically, a model displaying ANM comprises an underdamped Brownian particle in a sinusoidal potential driven by an unbiased periodic force $f(t)=A\sin(\omega t)$ \cite{speer2007transient}:
\begin{equation}
    m\ddot{x}(t)+\dot{x}(t)=-U'(x)+F+f(t)+\sqrt{2D_T}\xi(t).
    \label{minimal_Speer}
\end{equation}
In 2007, Speer \textit{et al.} established a no-go theorem outlining the necessary conditions for the emergence of ANM in continuous-time dynamics. Crucially, Eq.~(\ref{minimal_Speer}) is ergodic for any finite thermal noise intensity $D_T$, ensuring that the stationary average velocity $\langle v\rangle$ is unique and entirely independent of the initial conditions. Since the static bias $F$ is the sole source of asymmetry in Eq.~(\ref{minimal_Speer}), given that the time-average of the periodic driving vanishes, $\langle f(t)\rangle=0$, the system satisfies the symmetry property that the transformation $F\to-F$ results in $\langle v\rangle\to-\langle v\rangle$. Consequently, directed transport vanishes in the absence of a bias, i.e., $\langle v\rangle=0$ for $F=0$. To realize ANM, we seek regimes where the external force $F$ and the resulting average velocity $\langle v\rangle$ possess opposite signs.

In a purely linear system, according to Newton's second law, switching from $F=0$ to a finite force necessarily induces an acceleration in the direction of that force, leading to a velocity with the same sign. In Eq.~(\ref{minimal_Speer}), however, this straightforward intuition can break down due to the interplay of multiple forces; specifically, the periodic force profile $-U'(x)$ introduces essential nonlinearities into the equations of motion. Nonlinearity is thus a fundamental prerequisite for the occurrence of ANM.

At first glance, a steady-state velocity $\langle v\rangle$ directed opposite to a constant force $F$ appears to violate Le Chatelier's principle, thermodynamic stability criteria, and the second law of thermodynamics. Such equilibrium constraints, however, do not apply to systems driven far from equilibrium~ \cite{zia2002getting}. Accordingly, the second essential condition for ANM is that the system must operate under nonequilibrium conditions. In Eq.~(\ref{minimal_Speer}), this requirement is fulfilled by the time-dependent periodic driving $f(t)$.

To see why inertia is crucial, let us consider a general system described by the overdamped Langevin equation:
\begin{equation}
    \dot{x}(t)=h(x(t),t)+F+g(x(t),t)\xi(t),
    \label{model_nogo}
\end{equation}
where $h(x(t),t)$ and $g(x(t),t)$ are continuous functions. Let $x_1(t)$ and $x_2(t)$ be two distinct trajectories satisfying Eq.~(\ref{model_nogo}) for identical initial conditions, $x_1(0)=x_2(0)$, and identical realizations of the noise $\xi(t)$, differing only in their static biases $F_1$ and $F_2$. Assuming $F_1>F_2$, at any time $t$ where the trajectories intersect ($x_1(t)=x_2(t)$), Eq.~(\ref{model_nogo}) implies that $\dot{x}_1(t)-\dot{x}_2(t)=F_1-F_2>0$. By continuity, this guarantees that $x_1(t)\ge x_2(t)$ for all $t\ge0$, meaning that the average velocity $\langle v\rangle$ must be a monotonically increasing function of the bias $F$. The resulting no-go theorem dictates that if $\langle v\rangle=0$ at $F=0$, the force and velocity can never have opposite signs in purely overdamped dynamics. Consequently, the inertial term $m\ddot{x}(t)$ represents a third necessary condition for the emergence of ANM in this class of systems. 

Conversely, completely omitting the dissipation term ($\gamma\to0$ while keeping $\gamma k_B T$ fixed) corresponds to an infinitely high effective temperature, which completely suppresses the influence of the periodic potential $-U'(x)$, thereby precluding ANM once again.
\subsection{Ratchet Effect}
Finally, the \textit{ratchet effect} refers to the generation of directed transport in the complete absence of a net external bias~ \cite{reimann2002brownian,hanggi2009artificial}, i.e., 
\begin{equation}
    \langle v\rangle\Big|_{F=0}\ne0.
\end{equation}
This phenomenon allows one to extract directed motion from random fluctuations and perform useful work against an external load. At thermal equilibrium, such a mechanism would constitute a perpetuum mobile of the second kind; hence, the ratchet effect can only emerge when the system is driven out of thermodynamic equilibrium.

\begin{figure}
    \centering
    \includegraphics[width=0.49\textwidth]{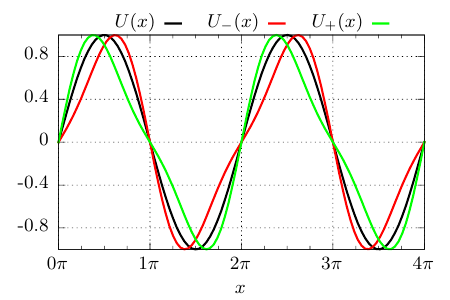}
    \caption{Spatially symmetric sine potential $U(x)$ and ratchet potentials $U_-(x)$ and $U_+(x)$.}
    \label{fig:potential}
\end{figure}

The emergence of directed transport is fundamentally connected to Curie's principle, which dictates that the symmetries of the causes must be reflected in their effects~ \cite{reimann2002brownian,curie1894symetrie,}. Because the external constant force vanishes ($F=0$), some other component of the system must break the spatial inversion symmetry. For instance, the symmetric sinusoidal potential $U(x)=\sin(x)$ can be replaced by asymmetric sawtooth or ratchet potentials, such as:
\begin{equation}
    U_{\pm}(x)=0.908(\sin(x)\pm0.25\sin(2x))
\end{equation}
presented in Fig. \ref{fig:potential}. The scaling factor $0.908$ ensures that these asymmetric profiles remain directly comparable to the standard sinusoidal potential, preserving identical barrier heights ($\Delta U=\Delta U_{\pm}=2$) and spatial periods ($L_U=L_{U_\pm}=2\pi$). As emphasized, being out of equilibrium is a strict requirement for the ratchet effect~ \cite{reimann2002brownian,hanggi2009artificial}; in an equilibrium state, detailed balance prevents any net transport. This seemingly paradoxical behavior was originally conceptualized by Smoluchowski~ \cite{von1912experimental} and famously analyzed by Feynman~ \cite{feynman2013lectures}. 

Depending on how the system is driven out of equilibrium with time dependent function $f(t)$, we distinguish two main classes of ratchets, i.e. \textit{pulsating} and \textit{tilting} ratchets \cite{reimann2002brownian}. In the former, the time dependent function $f(t)$ appears in the periodic potential, i.e. after the following transformation $U(x)\to U(x,f(t))$. When the it is of the form
\begin{equation}
 U(x,f(t))=U(x)(1+f(t)),
\end{equation} the system is called fluctuating potential ratchet. On the other hand, we call it traveling potential ratchet if
\begin{equation}
U(x,f(t))=U(x-f(t)).
\end{equation} When $f(t)$ is periodic function or stationary process, the potential $U(x,f(t))$ is called genuine traveling potential, because no systematic long time drift $u:=\lim_{t\to\infty}f(t)/t$ is generated by $f(t)$. Potentials in tilting ratchets are time independent and $f(t)$ is an additive force like in Eq. (\ref{minimal_Speer}). When $f(t)$ is a stochastic process, the system is called the fluctuating force ratchet and when $f(t)$ is a periodic function, the rocking ratchet \cite{bartussek1994periodically}. Besides these two main classes of ratchet systems there are two those that do not have time dependent potential or additional additive force. In Seebeck and temperature ratchets the temperature is space or time dependent respectively \cite{ashcroft1976solid,reimann1996brownian}. Please note, that these classifications correspond to minimal models. The system may be both pulsating and tilting ratchet simultaneously.

Ratchet systems are particularly useful for particle sorting  \cite{mcfaul2012cell,van1999brownian,hu2017brownian}. They appear in wide range of physical systems, from molecular motors \cite{astumian1997thermodynamics}, optical systems \cite{faucheux1995optical}, dielectric potentials  \cite{rousselet1994directional}, superconductors  \cite{villegas2003superconducting} or even macroscopic Leidenfrost droplets on sawtooth surface \cite{linke2006self}. For a more comprehensive overview of the ratchet effect and its various implementations, we refer the reader to the seminal review by Reimann~ \cite{reimann2002brownian}.

\section{Giant Transport Enhancement}
\label{GiantEnhancement_chapt}
\subsection{Overdamped Dynamics}

The primary objective of this thesis is to investigate transport phenomena driven by active fluctuations. We begin by considering the simplest case of overdamped dynamics, examining a dimensionless formulation of Eq.~(\ref{const_pot}) wherein the constant force $F$ is replaced by white Poisson shot noise $\eta(t)$ characterized by a mean spiking rate $\lambda$, an average amplitude $\zeta$, and an amplitude distribution $\rho(z)$. This Langevin equation takes the form:
\begin{equation}
    \dot{x}(t)=-\varepsilon U'(x)+\eta(t)+\sqrt{2D_T}\xi(t),
    \label{overdamped_poisson}
\end{equation}
where $D_T=0.01$ denotes the intensity of the thermal fluctuations. The parameter $\varepsilon$ modulates the barrier height of the spatial potential $U(x)=\sin(x)$ and corresponds to half of the total barrier height of the potential $\varepsilon U(x)$. For further details regarding the scaling procedure, we refer the reader to Appendix~A in Paper~\hyperref[A1]{A1}.

Alternatively, the statistical behavior of the system can be described by a Fokker-Planck framework, wherein the probability density function $P(x,t)$ of the state variable $x(t)$ satisfies the following integro-differential master equation~ \cite{hanggi1978derivations,hanggi1980langevin}:
\begin{equation}
    \frac{\partial}{\partial t}P(x,t)=\varepsilon\frac{\partial}{\partial x}[ U'(x)P(x,t)]+D_T\frac{\partial^2}{\partial x^2}P(x,t)+\lambda\int_{-\infty}^{\infty}[P(x-z,t)-P(x,t)]\rho(z)\,\mathrm{d}z.
    \label{FP_overdamped}
\end{equation}
As this equation cannot be solved analytically in the general case, the results presented herein are primarily obtained via high-precision numerical simulations executed on graphics processing units (GPUs)~ \cite{spiechowicz2015gpu}. Employing a Monte Carlo integration scheme~ \cite{kim2007numerical,platen2010numerical}, hundreds of thousands of independent realizations of Eq.~(\ref{overdamped_poisson}) are simulated concurrently to perform the ensemble averaging. 

\begin{figure}
    \centering
    \includegraphics[width=0.49\linewidth]{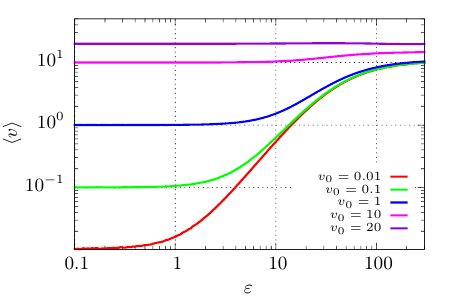}    
    \includegraphics[width=0.49\linewidth]{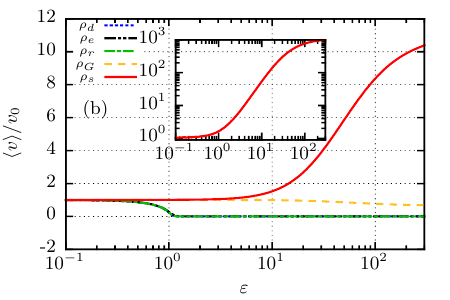}
    \caption{Average velocity $\langle v\rangle$ as a function of the barrier height $\varepsilon$ (left panel, \hyperref[A1]{A1}) and the corresponding rescaled velocity $\langle v\rangle/v_0$ across various noise amplitude distributions $\rho(z)$ (right panel, \hyperref[A2]{A2}). The inset highlights the giant velocity enhancement ($\langle v\rangle/v_0 \gg 1$) achieved under a skew-normal distribution $\rho_s(z)$ for a small free-particle velocity $v_0 = 0.01$.}
    \label{fig:giant}
\end{figure}

In the absence of a periodic potential ($\varepsilon=0$), the average velocity of a free Brownian particle, denoted by $v_0$, is identical to the net average bias analogous to a deterministic system driven by a constant force $F$:
\begin{equation}
    \langle v\rangle\big|_{\varepsilon=0}=\langle \eta(t)\rangle=\lambda \zeta=v_0.
\end{equation}
Intuitively, increasing the potential barrier $\varepsilon$ should hinder directed transport, leading to a reduced velocity $\langle v\rangle < v_0$. However, we find that in the presence of active fluctuations $\eta(t)$ governed by a properly tuned skew-normal amplitude distribution $\rho(z)$, the counter-intuitive phenomenon of giant transport enhancement can emerge. As illustrated in the left panel of Fig.~\ref{fig:giant}, the average velocity increases monotonically with the barrier height $\varepsilon$ before reaching a plateau in the limit of infinitely high barriers, $\varepsilon \to \infty$. When quantifying this effect via the rescaled velocity $\langle v\rangle/v_0$, the enhancement can span several orders of magnitude, particularly in regimes of weak average bias.

A precise tuning of the amplitude distribution $\rho(z)$ with respect to the profile of the spatial potential is a strict prerequisite for the emergence of this velocity enhancement. In Paper~\hyperref[A2]{A2}, we systematically investigated the specific statistical characteristics of the noise amplitude required to induce this phenomenon. By comparing the dynamics across all amplitude distributions detailed in Chapter~3.2, namely deterministic, exponential, Erlang, Gaussian, and skew-normal statistics, we demonstrate in the right panel of Fig.~\ref{fig:giant} that transport enhancement is uniquely supported by the skew-normal distribution. Specifically, the amplitude distribution must satisfy a stringent set of criteria: it must be bidirectional (permitting both positive and negative fluctuations), possess a small average amplitude $\zeta$, feature a variance $\sigma^2$ that can be varied independently of the mean $\zeta$, and exhibit broken inversion symmetry (characterized by a non-vanishing skewness $\chi \neq 0$). Among the considered models, only the skew-normal distribution satisfies all of these necessary conditions simultaneously.

\begin{figure}
    \centering
    \includegraphics[width=0.49\linewidth]{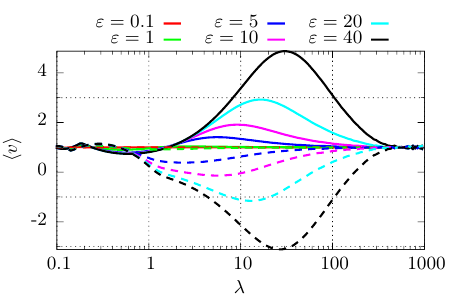}
    \includegraphics[width=0.49\linewidth]{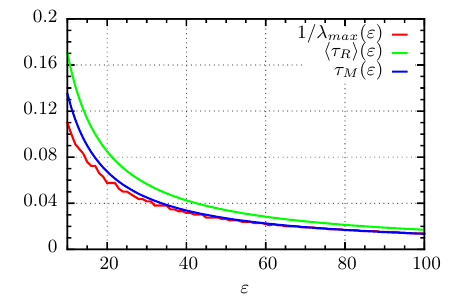}
\caption{Average velocity $\langle v\rangle$ vs. mean spiking rate $\lambda$ with a fixed average bias $\langle \eta(t)\rangle=1$ for different barrier heights $\varepsilon$ (left panel \hyperref[A1]{A1}), and the dependence of the optimal mean spiking rate $\lambda_{max}(\varepsilon)$, average relaxation time $\langle \tau_R\rangle(\varepsilon)$, and median relaxation time $\tau_M(\varepsilon)$ on the barrier height $\varepsilon$ (right panel \hyperref[A2]{A2}). The solid and dashed lines in the left panel correspond to positive skewness ($\chi=0.99$) and negative skewness ($\chi=-0.99$), respectively.}
    \label{fig:time}
\end{figure}

In the pursuit of optimal transport (the highest average velocity $\langle v\rangle$ for a given average bias $\langle \eta(t)\rangle$), we investigated the dependence of the average velocity $\langle v\rangle$ on the noise parameters, i.e., the mean spiking rate $\lambda$ and the average amplitude $\zeta$, with a fixed average bias $\langle\eta(t)\rangle=\lambda\zeta$. The results are shown in the left panel of Fig. \ref{fig:time}. The velocity maximum $v_{max}$ appears at a higher mean spiking rate $\lambda$ when the barrier height $\varepsilon$ is increased. The inverse of the mean spiking rate $\lambda$ is the average time between impulses, $\tau_P=1/\lambda$, which decreases as $\lambda$ increases.

The relaxation time $\tau_R$, which is the time it takes for the particle to reach the potential minimum, is inversely proportional to the barrier height, $\tau_R \propto 1/\varepsilon$; that is, for steeper potentials, the minimum is reached faster. In the right panel of Fig. \ref{fig:time}, we compare the average relaxation time $\langle\tau_R\rangle(\varepsilon)$ or the median relaxation time $\tau_M(\varepsilon)$ for a given barrier height $\varepsilon$ with the average time between $\delta$-impulses, calculated from the optimal mean spiking rate $\lambda_{max}$ for which the velocity is maximal. These quantities are approximately equal, particularly $\tau_M(\varepsilon)\approx 1/\lambda_{max}$. We can conclude that optimal transport occurs in the statistical resonance regime, where, on average, the Brownian particle is not affected by another active fluctuation until it reaches the vicinity of the potential minimum. 

\begin{figure}
    \centering
    \includegraphics[width=0.49\linewidth]{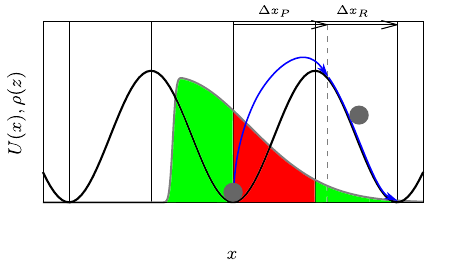}
    \includegraphics[width=0.49\linewidth]{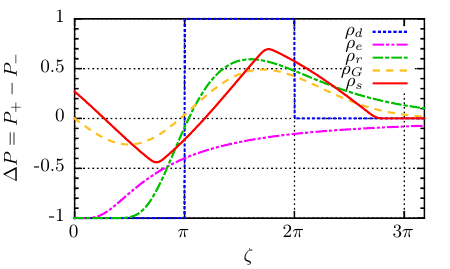}
    \caption{Schematic representation of the jump-relaxation process as a phenomenological description of the system dynamics in the absence of thermal fluctuations, with an exemplary skewed amplitude distribution $\rho(z)$ optimal for transport enhancement (left panel \hyperref[A3]{A3}). Probability difference $\Delta P=P_+-P_-$ of landing on slopes with positive ($P_+$) and negative ($P_-$) relaxation $\Delta x_R$ (right panel \hyperref[A2]{A2}).}
    \label{fig:jump}
\end{figure}

To quantify the transport enhancement, we construct a toy model inspired by the statistical resonance condition for optimal transport. Since the role of thermal fluctuations $D_T=0.01$ in transport is negligible, we focus on two aspects of the dynamics: the active jump $\Delta x_P=z_i$ and the relaxation towards the potential minimum $\Delta x_R$. We set the initial position of the Brownian particle at the minimum and calculate the total displacement $\Delta x$ resulting from a single $\delta$-impulse $\Delta x_P$ and the subsequent relaxation down the potential slope $\Delta x_R$, as shown in Fig. \ref{fig:jump}. We call this toy model the jump-relaxation process. The average velocity is then
\begin{equation}
\langle v\rangle=\frac{\langle \Delta x\rangle}{\tau_P}=\lambda(\langle \Delta x_P+\Delta x_R\rangle)=\lambda(\zeta+\langle\Delta x_R\rangle).
\label{jr_velocity}
\end{equation}
The enhancement, i.e., the magnitude of the rescaled velocity $\langle v\rangle/v_0$, then reads
\begin{equation}
\frac{\langle v\rangle}{v_0}=\frac{\lambda\zeta+\lambda\langle\Delta x_R\rangle}{\lambda\zeta}=1+\frac{\langle\Delta x_R\rangle}{\zeta}.
\label{enhancement}
\end{equation}
For optimal transport, the average relaxation $\langle\Delta x_R\rangle$ must be maximal for a given average amplitude $\zeta$. Because the maximal relaxation distance $\max|\Delta x_R|$ is bounded by the geometry of the potential, the mean amplitude $\zeta$ must be small, i.e., $\zeta\to0$.

In the simplest version of the jump-relaxation process, the particle always reaches a minimum. This can be realized in the rare fluctuation limit $\lambda\to0$. Then, the displacement $\Delta x$ can only be a multiple of the potential period, $\Delta x=kL$, where $k\in\mathbb{Z}$. The probability of such a transport event can be calculated as
\begin{equation}
    p_{\pm kL}=\int_{-L/2\pm kL}^{L/2\pm kL}\rho(z)dz, \quad k\in\mathbb{Z},
    \label{probability_coarse}
\end{equation}
where we have assumed that the potential minimum is located exactly between two maxima. The average velocity then reads
\begin{equation}
    \langle v_p\rangle=L\lambda\sum_{i=1}^{\infty}k(p_{kL}-p_{-kL}).
    \label{coarse_grained_overdamped}
\end{equation}
This rough approximation only yields the average velocity corresponding to the plateaus in Fig. \ref{fig:giant} and is independent of the barrier height $\varepsilon$. The dependence on the mean spiking rate $\lambda$ is linear, which would imply that one could increase the impulse rate indefinitely and expect the transport to increase continuously. We know this is not true from the statistical resonance condition presented in Fig. \ref{fig:time}; therefore, a more detailed model is needed.

Before diving into a more complex version of the jump-relaxation model, let us consider only the relaxation from the four closest slopes of the periodic potential, i.e. regions
\begin{enumerate}[label=\Roman*]
\item $ x\in (-L,-L/2)$,
\item $x\in (-L/2,0)$,
\item $x \in (0, L/2)$,
\item $x\in (L/2, L)$
\end{enumerate}
 (the optimal skew-normal amplitude distribution $\rho_s(z)$ has a variance $\sigma^2$ of the same order as the length of the potential period, $\sigma^2\approx L$, so this approximation is sufficient). By integrating the amplitude distribution $\rho(z)$ over these intervals, we can obtain the probabilities of landing on these slopes (the green and red areas in the right panel of Fig. \ref{fig:jump}, respectively). Note that the relaxation $\Delta x_R$ in ranges I and III is negative, while the relaxation in ranges II and IV is positive. If we sum the probabilities of positive relaxations $P_+$ and subtract the corresponding probability of negative relaxation $P_-$,
\begin{equation}
P_+=\int_{-L/2}^{0}\rho(z)dz +\int_{L/2}^{L}\rho(z)dz,
\end{equation}
\begin{equation}
P_-=\int_{-L}^{-L/2}\rho(z)dz +\int_{0}^{L/2}\rho(z)dz,
\end{equation} 
we can gain insight into the necessary characteristics of the amplitude distribution for enhancement to occur, as shown in the right panel of Fig. \ref{fig:jump}. Only the skew-normal distribution allows for a positive average relaxation in the small-amplitude limit $\zeta\to0$, where the giant enhancement occurs.

To obtain an accurate estimate of the average velocity, we relax the condition of the particle's initial position being strictly at the minimum to a random position $x$ within the potential period ($x\in[-L/2, L/2]$), represented by the position probability density function $p(x)$. The relaxation $\Delta x_R$ no longer brings the particle all the way to the potential minimum. Now, $\Delta x_R$ corresponds to the relaxation down the potential slope from the position $x+z$ during a time interval $\tau$ between consecutive $\delta$-impulses; that is, the function $\Delta x_R(x+z,\tau)$ depends on the geometry of the potential. For sinusoidal and piecewise-linear periodic potentials, it can even be derived analytically, see articles \hyperref[A2]{A2} and \hyperref[A6]{A6}. The time $\tau$ represents the time interval between consecutive fluctuations, which for white Poisson noise is exponentially distributed:
\begin{equation}
    \phi(\tau)=\theta(\tau)\lambda\exp(-\lambda\tau),
\end{equation}
where $\theta(\tau)$ is the Heaviside step function. The average relaxation distance $\langle \Delta x_R\rangle$ is obtained by averaging over all three probability distributions:
\begin{equation}
    \langle \Delta x_R\rangle=\int_{-L/2}^{L/2}\int_{-\infty}^{\infty}\int_{0}^{\infty}\phi(\tau)\rho(z)p(x) \Delta x_R(x+z,\tau)d\tau dzdx.
    \label{full_jr}
\end{equation}
In the rare impulse limit $\lambda\to0$, the initial probability density function tends to the Dirac delta function, $p(x)|_{\lambda\to0}\approx\delta(x)$. In addition, the dependence of $\Delta x_R(x+z,\tau)$ on the time $\tau$ disappears, and the average relaxation distance depends solely on the amplitude distribution $\rho(z)$, i.e., $\Delta x_R(x+z,\tau)\to\Delta x_R(z)$. Consequently, the detailed model converges to the previously described coarse-grained model:
\begin{equation}
    \langle v_p\rangle=\lambda(\zeta+\langle \Delta x_R\rangle|_{\lambda\to0}).
\end{equation}

\begin{figure}
    \centering
    \includegraphics[width=0.49\linewidth]{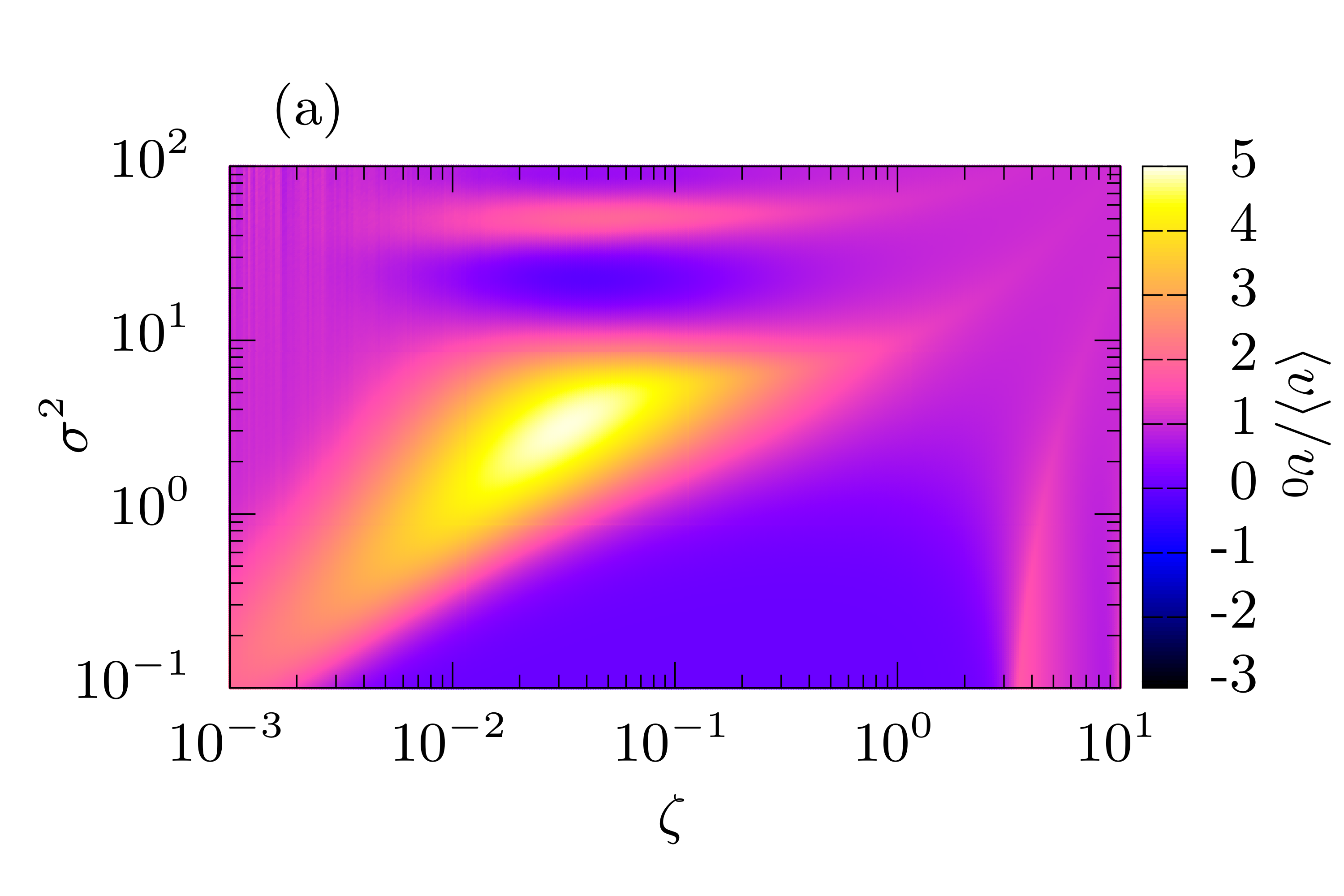}
    \includegraphics[width=0.49\linewidth]{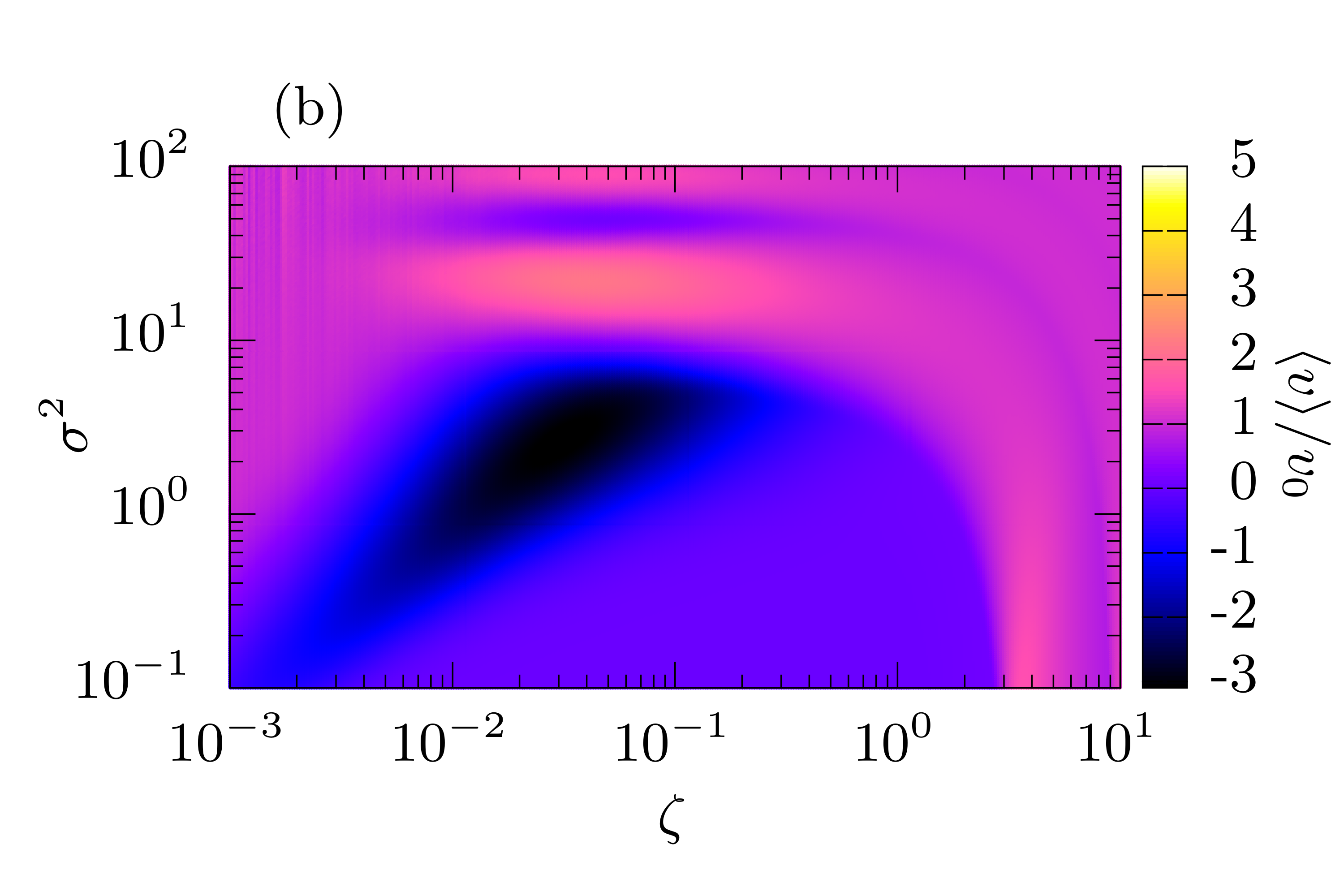}
    \caption{The rescaled velocity $\langle v\rangle/v_0$ (color-coded) as a function of the mean amplitude $\zeta$ and variance $\sigma^2$ of active fluctuations $\eta(t)$ for positive $\chi=0.99$ and negative $\chi=-0.99$ skewness in panels $(a)$ and $(b)$, respectively. The average bias value is fixed at $\langle\eta(t)\rangle=\lambda\zeta=1$. Figure reproduced from \hyperref[A3]{A3}.}
    \label{fig:mapa_parametry}
\end{figure}

In \hyperref[A3]{A3}, we studied the influence of all the statistical properties of the amplitude distribution $\rho(z)$ on the directed velocity. In Fig. \ref{fig:mapa_parametry}, we present a comparison of the rescaled velocities $\langle v\rangle/v_0$ mapped against the average amplitude $\zeta$ and variance $\sigma^2$. There, two types of decaying oscillations can be observed. The first type occurs when the variance is very small, $\sigma^2\to0$, and the average amplitude $\zeta$ increases. In this case, the situation is trivial, as the skew-normal distribution behaves similarly to a deterministic one, $\Delta x_R(x+z,\tau)\approx \Delta x_R(x+\zeta,\tau)$. When $\zeta$ passes the potential maximum, the relaxation switches from the minimum, $\langle\Delta x_R\rangle\approx-L/2$, to the maximum, $\langle\Delta x_R\rangle\approx L/2$. To better illustrate this, let us consider the closest maximum at $x=L/2$:
\begin{equation}
    \lim_{x+\zeta\to{(L/2)^-}}\Delta x_R(x+\zeta,\tau)=-\frac{L}{2},
\end{equation}
\begin{equation}
    \lim_{x+\zeta\to{(L/2)^+}}\Delta x_R(x+\zeta,\tau)=\frac{L}{2},
\end{equation}
This oscillation vanishes because $\langle\Delta x_R\rangle$ is bounded, whereas $\zeta$ is unbounded in the Eq. (\ref{enhancement}).

The second type is much more physically relevant; it occurs for a small but finite $\zeta$ (the limit $\zeta\to0$ implies $\lambda\to\infty$, which breaks the statistical resonance as shown in Fig. \ref{fig:time}, since the $\delta$-impulses occur too frequently to allow the particle to relax toward the vicinity of the potential minimum). As the variance $\sigma^2$ increases, the particle can jump into more distant potential wells ($L+nL$), and the average relaxation $\langle \Delta x_R\rangle$ can change its sign. Since the average amplitude $\zeta$ is fixed, the decay of these oscillations is not as rapid as it was for the first type. In this regime, the decay is caused by the ``flattening'' of the amplitude distribution for larger variances $\sigma^2$; i.e., the contrast between positive and negative relaxations disappears.

The scientific significance of the findings presented in this chapter is highlighted first by their biological relevance, as this mechanism helps explain how biological systems, such as molecular motors (e.g., dynein and kinesin) traveling along spatially periodic structures like microtubules, achieve highly effective intracellular transport \cite{ezber2020dynein,ariga2021noise}. Additionally, these results offer crucial insights for technological applications, providing a framework for designing efficient, biologically inspired micro- and nanoscale machines that can successfully harness active noise to optimize transport. Finally, we underscore the viability of these concepts through experimental corroboration, suggesting that these transport results can be directly validated using real-world setups like colloidal particles in optically generated periodic potentials \cite{park2020rapid,paneru2021transport}.

\subsection{Inertial Dynamics}

\begin{figure}
    \centering
    \includegraphics[width=0.49\linewidth]{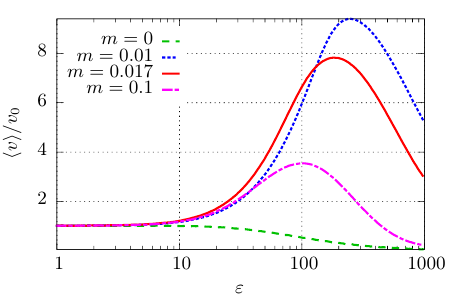}
    \caption{Dependence of the rescaled velocity $\langle v\rangle/v_0$ on the barrier height $\varepsilon$ for different values of inertia $m$. Figure reproduced from \hyperref[A4]{A4}.}
    \label{fig:inertial_eps}
\end{figure}

Large fluctuations in the overdamped limit may cast doubt on the validity of the studied transport mechanism. In the first-order stochastic differential Langevin equation \eqref{overdamped_poisson}, a Poisson kick results in an instantaneous particle displacement equal to the amplitude of the fluctuation, thereby ignoring the potential barrier. Therefore, the corresponding inertial system has also been investigated in \hyperref[A4]{A4} and \hyperref[A5]{A5}. 

Active matter has historically been modeled primarily within the overdamped limit because it serves as a prototype for biological systems operating in highly viscous environments \cite{ten2011brownian,volpe2014simulation}. Recently, however, inertial dynamics have rapidly gained attention in the physics community \cite{lowen2020inertial,caprini2021inertial}. In particular, we focus on the interface between overdamped and underdamped limits, where the inertia is small.
The inertial dynamics of our system are described by the following dimensionless Langevin equation:
\begin{equation}
m\ddot{x}(t)+\dot{x}(t)=-U'(x)+\eta(t)+\sqrt{2D_T}\xi(t).
    \label{inertial_poisson}
\end{equation}
In the limit $m\to0$, Eq. (\ref{overdamped_poisson}) is recovered. As shown in Fig. \ref{fig:inertial_eps}, transport disappears for a large barrier height $\varepsilon\to\infty$, contrary to the overdamped limit where the average velocity $\langle v\rangle$ approaches the plateau shown in Fig. \ref{fig:giant}. This behavior is related to the microscopic dynamics of how the particle responds to a single active fluctuation with amplitude $z_i$. In the overdamped limit, the discontinuity occurs in position space, i.e., $\Delta x=z_i$; on the other hand, in inertial dynamics, it appears in momentum space as $\Delta p=z_i$, which results in a velocity change of $\Delta v=z_i/m$. 

In overdamped dynamics, the instantaneous velocity $v(t)$ and kinetic energy cannot be defined; barrier crossing does not depend on them, but rather on moving against the potential gradient, which is ignored during the instantaneous displacement caused by large $\delta$-impulses. In inertial dynamics, to cross a potential barrier, the fluctuation must provide sufficient energy to cover both the distance toward the potential maximum and the difference in potential energy between the maximum and the initial position of the particle. If the fluctuation kicks the particle exactly at the potential minimum, this required potential energy gain is $\Delta E_p=2\varepsilon$. As the barrier height $\varepsilon$ increases, transport over the barrier eventually becomes impossible. 

Transport enhancement can still be achieved in inertial dynamics, as shown in Fig. \ref{fig:inertial_eps}. However, it is most effective in the strong-damping regime. Larger inertia destabilizes the statistical resonance introduced in the previous chapter, where the average time between fluctuations corresponded to the average relaxation time toward the potential minimum. In the inertial system, the motion resulting from a fluctuation takes proportionally longer to occur as inertia increases. To illustrate this, let us consider a single deterministic free particle obeying the following equation of motion:
\begin{equation}
    m \ddot{x}(t)=-\dot{x}.
\end{equation}
Solving this ordinary differential equation for the initial position $x(0)=0$ and initial velocity $\dot{x}(0)=z_i/m$ yields the solution:
\begin{equation}
    x(t)=z_i \left(1- \exp\left(-\frac{t}{m}\right)\right),
\end{equation}
\begin{equation}
    \dot{x}(t)=\frac{z_i}{m}\exp\left(-\frac{t}{m}\right).
\end{equation}
As can be seen, the particle approaches the position $x=z_i$ faster for a smaller mass $m$.

\begin{figure}
    \centering
    \includegraphics[width=0.49\linewidth]{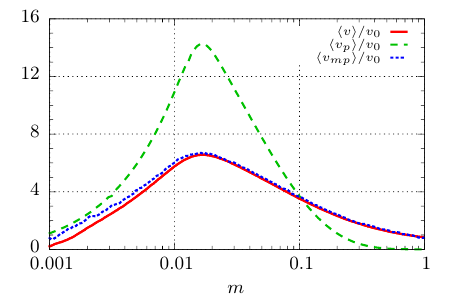}
    \includegraphics[width=0.49\linewidth]{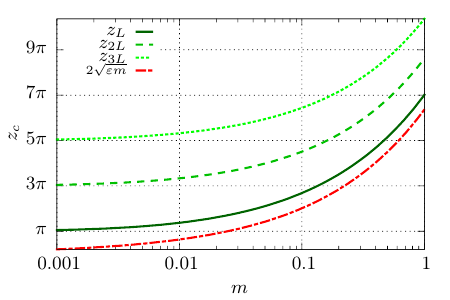}
    \caption{Rescaled velocity $\langle v\rangle/v_0$ as a function of inertia $m$ for the simulated data and different approximations (left panel), and the critical barrier-crossing fluctuation amplitude $z_c$ as a function of mass $m$ (right panel). Figure reproduced from \hyperref[A4]{A4}.}
    \label{fig:inertial_mass}
\end{figure}

For carefully chosen amplitude statistics, the enhancement may appear only within a specific range of inertia $m$, as shown in Fig. \ref{fig:inertial_mass}. Here, the variance $\sigma^2$ of the amplitude distribution is chosen such that transport vanishes ($\langle v\rangle\approx0$) in the overdamped dynamics. Therefore, the average directed velocity is zero in the limit of vanishing inertia, $m\to0$. In the strong-damping regime, however, where the mass $m$ has a small but finite value, transport enhancement is detected. Therefore, for this particular noise statistics, this enhancement can be considered inertia-induced. For large inertia $m$, it disappears again because, in this mass regime, the localization of the particle at the potential minimum while awaiting the next fluctuation is significantly weakened. This breaks the statistical resonance, as active fluctuations no longer act on the particle at the same average position (vicinity of the potential minimum). The presence of a velocity maximum at an intermediate mass $m_0$ provides opportunities for potential applications, such as the sorting of particles based on their mass $m$.

Because the jump cannot be cleanly separated from the relaxation in inertial dynamics, we must adapt our toy model. The coarse-grained approximation in Eq. (\ref{coarse_grained_overdamped}) can be easily adapted by changing the critical barrier-crossing fluctuation amplitude $z_c$. Instead of using the distances between potential minima and the corresponding potential maxima, $L/2+nL$ ($n\in \mathbb{Z}$), we use the critical amplitudes presented in the right panel of Fig. \ref{fig:inertial_mass}. The red line there represents the dissipationless Hamiltonian limit, where all kinetic energy is transformed into potential energy, 
\begin{equation}
\frac{mv_c^2}{2}=\frac{z_c^2}{2m}=2\varepsilon,
\end{equation}
 yielding a critical amplitude value of 
\begin{equation}
z_c=2\sqrt{\varepsilon m}.
\end{equation} The critical amplitudes are approximately 
\begin{equation}
z_{kL}\approx kL+2\sqrt{\varepsilon m}.
\end{equation}
 For larger inertia $m$, the precisely calculated value of $z_c$ shown in Fig. \ref{fig:inertial_mass} differs from this expression because the deceleration is slowed down, with $\ddot{x}(t)\propto1/m$. The transfer probabilities to the $\pm k$-th potential period, $p_{\pm kL}$, and the average velocity, $\langle v_p\rangle$, then read:
\begin{equation}
    p_{\pm kL}=\pm \int_{\pm z_{kL}}^{\pm z_{(k+1)L}}\rho(z)dz, \quad k\in \mathbb{N} \setminus0.
    \label{inertial_probabilities}
\end{equation}
\begin{equation}
    \langle v_p\rangle=L\lambda\sum_{i=1}k(p_{kL}-p_{-kL}).
    \label{inertial_coursegrained}
\end{equation}
This approximation correctly predicts the optimal mass $m_o$ for which the rescaled velocity is maximized in Fig. \ref{fig:inertial_mass}. However, it severely overestimates the velocity. It suffers from the same problem as the coarse-grained model in the overdamped limit; namely, it assumes that the Brownian particle awaits the next fluctuation exactly at the potential minimum, which is valid only in the rare impulse regime, $\lambda\to0$. 

We adapt the transfer probabilities method to account for these limitations. The average transfer probabilities $p_\pm$ can be obtained from the simulated system trajectories by counting the barrier crossings $n_\pm$ and normalizing them by the expected number of $\delta$-impulses across $N$ simulations of duration $\mathsf{T}$, i.e.,
\begin{equation}
    p_{  \pm  }=\frac{1}{N}\frac{n_{\pm}}{\lambda \mathsf{T}}
\end{equation}
Since this method cannot distinguish between crossing a single barrier and crossing multiple barriers, Eq. (\ref{inertial_coursegrained}) transforms into:
\begin{equation}
    \langle v_{mp} \rangle=L\lambda(p_+-p_-).
\end{equation}
The velocity $\langle v_{mp}\rangle$ corresponds to $\langle v\rangle$ much better than the coarse-grained approximation $\langle v_p\rangle$.

\begin{figure}
    \centering
    \includegraphics[width=0.49\linewidth]{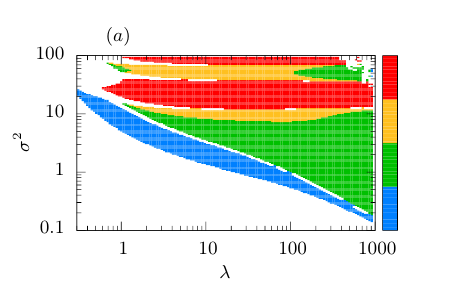}
    \includegraphics[width=0.49\linewidth]{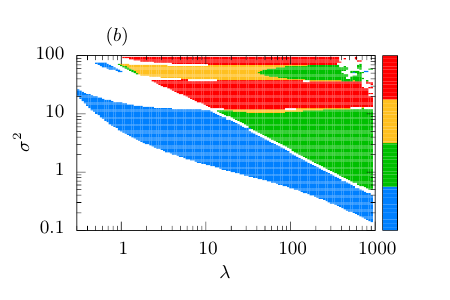}

    \caption{Comparison of the influence of inertia on transport enhancement regimes (strengthening \textcolor{Strengthening}{$\blacksquare$} $v_m > v_\gamma > v_0$, weakening \textcolor{Weakening}{$\blacksquare$} $v_0 < v_m < v_\gamma$, destructive \textcolor{Destructive}{$\blacksquare$} $v_m < v_0 < v_\gamma$, or constructive (inertia-induced) \textcolor{Constructive}{$\blacksquare$} $v_\gamma < v_0 < v_m$) for different inertia values, $m=0.01$ and $m=0.1$, in the left and right panels, respectively. Figure reproduced from \hyperref[A5]{A5}.}
    \label{fig:inertial_map}
\end{figure}

In \hyperref[A5]{A5}, we analyze in depth how the transport enhancement in the inertial system differs from that in the overdamped system presented in the previous chapter. Let $v_0$, $v_\gamma$, and $v_m$ denote the average velocities of a free Brownian particle, a particle in the overdamped system, and a particle in the inertial system, respectively. Concerning the studied mechanism, four distinct regimes can be identified. The influence of inertia can be positive; that is, a nonzero mass $m$ either \textit{strengthens} the enhancement ($v_0<v_\gamma<v_m$) or even \textit{constructs} it in a parameter regime where it was absent in the overdamped system ($v_\gamma<v_0<v_m$). Analogously, inertia may \textit{weaken} ($v_0<v_m<v_\gamma$) or completely \textit{destroy} the transport enhancement ($v_m<v_0<v_\gamma$). We refer to these regimes as strengthening, constructive, weakening, and destructive, respectively. Exemplary maps with the mean spiking rate $\lambda$ and variance $\sigma^2$ as coordinates for different values of inertia, $m=0.01$ and $m=0.1$, are shown in Fig.~\ref{fig:inertial_map}.

For most parameter sets, the influence of the mass $m$ is negative, especially for small variances $\sigma^2<10$, where it is either destructive or weakening. The parameter region corresponding to the destructive regime expands as inertia increases. When $\sigma^2>10$, the behavior becomes more interesting because the strengthening and constructive regimes appear at the edge of the enhancement regime and the transport-impeding ``gap'', respectively.

\begin{figure}
    \centering
    \includegraphics[width=0.49\linewidth]{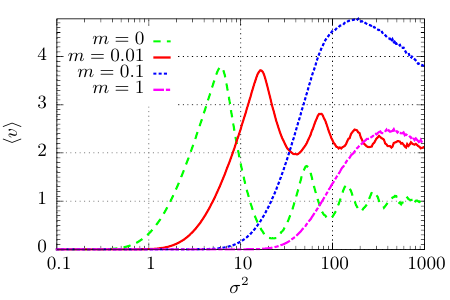}
    \caption{Dependence of the average velocity $\langle v\rangle$ on the variance of the amplitude distribution for different values of inertia $m$. Figure reproduced from \hyperref[A5]{A5}.}
    \label{fig:inertial_var}
\end{figure}

A key insight from Fig. \ref{fig:inertial_map} is that, for a fixed spiking rate $\lambda$, the entire range of inertial influences (i.e., strengthening, weakening, constructive, or destructive) appears across different amplitude variances $\sigma^2$. Accordingly, the variance acts as the definitive control parameter for the transport enhancement mechanism. The dependence of the average velocity on the variance is shown in Fig. \ref{fig:inertial_var}. The overdamped dynamics ($m=0$) serve as the baseline for comparison. 

Let us first consider the small-variance regime $\sigma^2<5$, spanning from the minimum to the first maximum of the average velocity $\langle v\rangle$ for the overdamped system ($m=0$). In this regime, a single active fluctuation can only induce transitions over the nearest potential barrier. Assuming the initial position is at the potential minimum, only positive fluctuations lead to directed transport, as negative fluctuations cannot provide sufficient energy. As inertia increases, the coupling between the potential period and the fluctuation amplitude changes, as shown in the right panel of Fig. \ref{fig:inertial_mass}; consequently, $\delta$-impulses capable of driving the particle over single or multiple potential barriers become less frequent. Therefore, the transfer probabilities defined in Eq. (\ref{inertial_probabilities}) decrease, leading to a negative, destructive, or weakening influence of inertia.

For larger variances $\sigma^2$, increasing the mass $m$ reduces the number of potential periods into which the particle can be kicked. For the positive skewness ($\chi=0.99$) focused on in this thesis, the tail of the distribution extends in the positive direction; therefore, transport in the negative direction is suppressed first. Since the difference between consecutive critical amplitudes $z_c$ remains approximately constant, $z_{(k+1)L}-z_{kL}\approx L$, the suppression of negative barrier crossings allows the Brownian particle to cross multiple potential barriers in the positive direction. This results in an average velocity $\langle v\rangle$ for $m=0.1$ that exceeds the maximum velocity observed in the overdamped dynamics ($m=0$).

As stated previously, the overdamped dynamics serve as a benchmark; for a given variance $\sigma^2>5$, the influence of inertia is initially positive. If $v_\gamma<v_0$, the influence is constructive, and if $v_\gamma>1$, it is strengthening. This positive influence persists until barrier crossings become unlikely due to a sufficient increase in the first critical amplitude $z_L$, at which point the influence of inertia becomes negative and transport vanishes ($\langle v\rangle\to0$).

This work advances the understanding of active matter by elucidating the fundamental and non-negligible role of inertia in transport processes driven by nonequilibrium fluctuations. Moving beyond the conventional overdamped approximation, which neglects particle mass, the results demonstrate that the strong-damping regime uncovers a versatile, inertia-induced mechanism capable of dynamically regulating the giant enhancement of directed transport. Rather than acting merely as a passive damping effect, inertia functions as a sensitive control parameter; it can uniquely induce transport amplification in parameter regimes where the effect is otherwise absent, and, depending on the variance of the active fluctuation amplitudes, it can strengthen, weaken, or completely suppress the phenomenon. By bridging the gap between simplified massless models and true inertial dynamics, these findings establish a comprehensive framework for mesoscopic physics, offering profound insights into both synthetic micro-machinery and biological systems, such as living cells where metabolically driven fluctuations inherently interact with inertial forces \cite{ezber2020dynein,ariga2021noise}.
\section{Absolute Negative Mobility}

\begin{figure}
    \centering
    \includegraphics[width=0.49\linewidth]{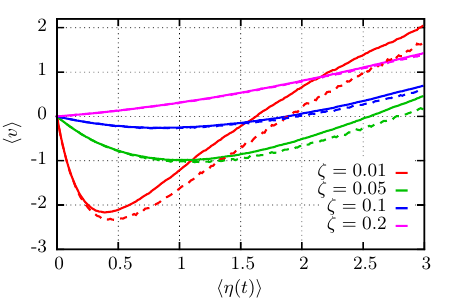}
    \includegraphics[width=0.49\linewidth]{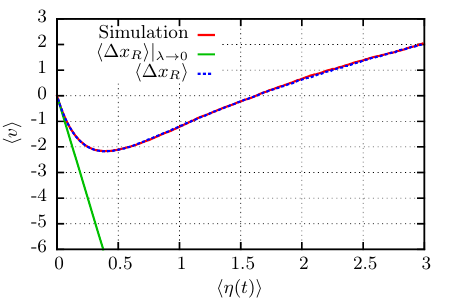}
    \caption{Dependence of the average velocity $\langle v\rangle$ on the average bias $\langle\eta(t)\rangle$ for a fixed average amplitude $\zeta$ and a varying mean spiking rate $\lambda$. In the left panel, the role of the average amplitude $\zeta$ is shown for a piecewise linear potential (solid lines) and a sine potential (dashed lines). In the right panel, the average velocities calculated with the jump-relaxation toy model $\lambda(\zeta+\langle \Delta x_R\rangle)$ are compared to the results from numerical simulations. Both the average relaxation in the rare impulse limit $\langle\Delta x_R\rangle|_{\lambda\to0}$ and that depending on the mean spiking rate $\langle\Delta x_R\rangle(\lambda)$ are included in the calculations of the average velocity $\langle v\rangle$. Figure reproduced from \hyperref[A6]{A6}.}
    \label{fig:ANM}
\end{figure}

One of the possible lines of inquiry regarding the studied system is how it responds to a small average perturbation. To do this, in \hyperref[A6]{A6} we examine how the average velocity $\langle v\rangle$ depends on the average bias $\langle \eta(t)\rangle$. Because the average stochastic force $\langle\eta(t)\rangle=\lambda\zeta$ depends on two parameters, the mean spiking rate $\lambda$ and the average amplitude $\zeta$, there are two distinct ways to achieve this: one can keep one parameter constant and vary the other.

If we assume that the average amplitude vanishes ($\zeta=0$), the active fluctuations vanish if the variance $\sigma^2$ of the amplitude distribution $\rho(z)$ depends on the average amplitude, as is the case for the exponential distribution $\rho_e(z)$. On the other hand, if the variance $\sigma^2$ is independent of the average amplitude $\zeta$, fluctuations always occur, and the system is out of equilibrium even when $\langle\eta(t)\rangle=0$.

Alternatively, if the mean spiking rate $\lambda$ is varied, the stochastic bias vanishes for all amplitude distributions $\rho(z)$ when $\lambda=0$. In the limit $\lambda\to0$, very rare fluctuations disturb an otherwise equilibrium system. 

Here, we restrict ourselves to positive average amplitudes $\zeta$, which implies that the average bias is non-negative ($\zeta>0 \implies \langle\eta(t)\rangle=\lambda\zeta\ge0$) since the mean spiking rate is non-negative ($\lambda\ge0$). Since the transport mechanism also appears in inertial dynamics, as shown in the previous chapter, we revert to the more minimal overdamped model described by Eq.~(\ref{overdamped_poisson}). By analyzing the parameter space, we find that in the linear response regime, anomalous transport in the form of absolute negative mobility (ANM) can be observed, as shown in Fig.~\ref{fig:ANM}. Because the Poisson noise is discontinuous, the no-go theorem forbidding ANM in the absence of inertia presented in Chapter~\ref{sec:Taxonomy} does not apply, and ANM appears even in overdamped dynamics.

Prior to our work, the minimal model for ANM required the fulfillment of four necessary conditions~ \cite{speer2007transient}. First, the system must be out of equilibrium and non-linear to allow transport in the direction opposite to the bias. Due to the continuity of ergodic dynamics, a smaller velocity for a larger bias is forbidden in a one-dimensional overdamped system with continuous noise. The final condition was non-zero damping of the system; in its absence, the role of the non-linear potential becomes negligible. Therefore, an inertial Brownian particle in a one-dimensional periodic sinusoidal potential driven out of equilibrium by an unbiased periodic force with symmetry broken by a constant deterministic force has been a paradigmatic minimal model of ANM for almost twenty years~ \cite{speer2007transient}. Our system is even more minimal: ANM emerges in overdamped dynamics rather than requiring inertia, the potential can be piecewise linear instead of a fully non-linear sinusoidal potential, and, in the absence of bias, the system is in equilibrium, whereas the previous minimal model remained out of equilibrium even without an external static force ($F=0$).

To establish when ANM can emerge, let us recall the expression for the average velocity in the jump-relaxation model, namely 
\begin{equation}
\langle v\rangle=\lambda(\zeta+\langle \Delta x_R\rangle).
\end{equation} It follows that for zero bias ($\langle\eta(t)\rangle=0$, corresponding to $\lambda=0$), the average velocity must also vanish ($\langle v\rangle=0$). Since the mean spiking rate is non-negative ($\lambda\ge0$), only a negative term inside the brackets $(\zeta+\langle \Delta x_R\rangle)$ can cause ANM. Therefore, ANM can only emerge when
\begin{equation}
  \mu_0=\frac{\langle v\rangle}{\langle \eta(t)\rangle} \Bigg|_{\lambda\to0}=1+\frac{\langle \Delta x_R\rangle|_{\lambda\to0}}{\zeta}<0 \iff  \langle\Delta x_R\rangle|_{\lambda\to0}<-\zeta.
    \label{condition_ANM}
\end{equation}

The average relaxation $\langle \Delta x_R\rangle$ depends non-linearly on the mean spiking rate $\lambda$; its magnitude $|\langle \Delta x_R\rangle|$ is a monotonically decreasing function of the mean spiking rate due to two factors: the shorter time interval between consecutive $\delta$-impulses and the detuning of the statistical resonance caused by the delocalization of the initial position. By continuity, the largest magnitude occurs in the rare fluctuation limit $\langle \Delta x_R\rangle|_{\lambda\to0}$, and the product $\lambda\langle \Delta x_R\rangle$ must exhibit an extremum at some intermediate mean spiking rate $\lambda$. As shown in Fig.~\ref{fig:ANM}, the linear dependence of the average velocity on the mean spiking rate quickly breaks down, meaning that the full jump-relaxation model from Eq.~(\ref{full_jr}) must be employed.

\begin{figure}
    \centering
    \includegraphics[width=0.49\linewidth]{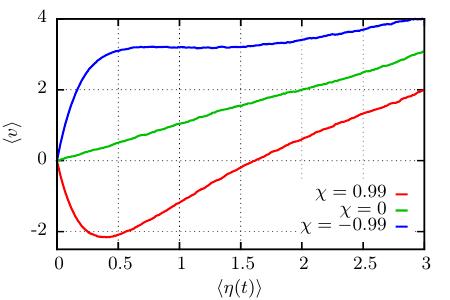}
    \includegraphics[width=0.49\linewidth]{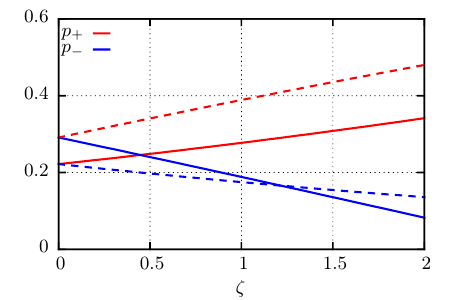}
    \caption{Dependence of the average velocity $\langle v\rangle$ on the average bias $\langle\eta(t)\rangle$ for a fixed average amplitude $\zeta$ and a varying mean spiking rate $\lambda$ for different values of skewness $\chi$ (left panel). Dependence of the transfer probabilities $p_\pm$ over the potential barriers in the positive ($+$) and negative ($-$) directions on the average amplitude $\zeta$ (right panel). The solid lines correspond to skewness $\chi=0.99$, while dashed ones to $\chi=-0.99$. Figure reproduced from \hyperref[A6]{A6}.}
    \label{fig:asymmetry}
\end{figure}

In the transport mechanism presented in this thesis, the asymmetry of the amplitude distribution plays a key role. As shown in Fig.~\ref{fig:asymmetry}, in the parameter regime considered in this chapter, a negative skewness $\chi=-0.99$ implies a positive average relaxation $\langle\Delta x_R\rangle>0$ and an enhancement of the average velocity compared to a free Brownian particle ($\langle v\rangle>v_0$). When the amplitude distribution is symmetric around its mean ($\chi=0$), this effect vanishes; since the average amplitude $\zeta=0.01$ is close to zero, the average velocity is comparable to that of a free Brownian particle under the same stochastic bias.

One might question whether the full jump-relaxation process is required to determine if ANM can be observed. For instance, one can naively try to calculate the probabilities of crossing the potential barriers in both directions, $p_{\pm}$,
\begin{equation}
    p_{\pm}=\pm\int_{\pm L/2}^{\pm \infty}\rho(z)dz
    \label{prob_pm}
\end{equation}
and check whether $p_- > p_+$. However, as shown in the right panel of Fig.~\ref{fig:asymmetry}, negative transfers are more probable even for average amplitudes $\zeta$ where ANM disappears, such as $\zeta=0.2$ in Fig.~\ref{fig:ANM}.


This chapter establishes a fundamental shift in the minimal requirements for absolute negative mobility (ANM), demonstrating that this anomalous transport phenomenon can robustly emerge within a purely overdamped, one-dimensional framework without requiring particle inertia, non-linear potentials, or non-stationary external driving. By utilizing a simple piecewise-linear symmetric periodic potential driven by active, non-Gaussian white Poisson shot noise, the presented model reveals that the interplay between impulsive active fluctuations and subsequent deterministic relaxation is entirely sufficient to produce a net negative steady-state velocity.

Beyond its theoretical contributions to nonequilibrium statistical physics, this minimal framework carries profound relevance for biological and microscale environments where viscous dissipation dominates and transport is fueled by active metabolic fluctuations rather than macroscopic gradients \cite{ezber2020dynein,ariga2021noise}. The structural simplicity of this model not only facilitates direct empirical validation via setups such as optical tweezers  \cite{park2020rapid,paneru2021transport} or Josephson junctions  \cite{nagel2008observation} but also provides a clear operational foundation for engineering highly selective particle separation devices that successfully exploit stochastic fluctuations as a functional transport resource.

\section{Ratchet Effect}

\begin{figure}[t]
    \centering
    \includegraphics[width=0.49\linewidth]{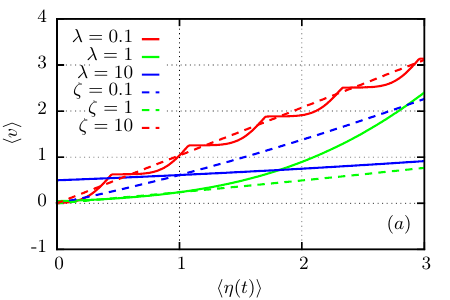}
    \includegraphics[width=0.49\linewidth]{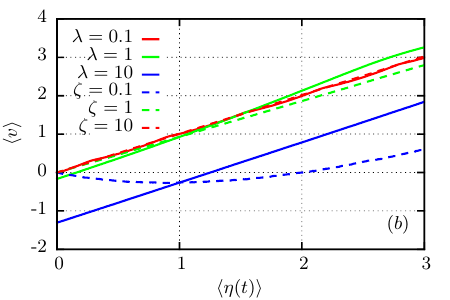}
    \caption{Average velocity $\langle v\rangle$ under different scaling schemes for the skew-normal amplitude distribution. Solid and dashed lines correspond to varying mean amplitude $\zeta$ and spiking rate $\lambda$, respectively. Panel (a) corresponds to variance $\sigma^2=1$, and panel (b) to $\sigma^2=20$.}
    \label{fig:ratchet}
\end{figure}

We next consider an alternative scaling of the mean active force, wherein the average amplitude $\zeta$ is varied at a fixed spiking rate $\lambda$. For an asymmetric amplitude distribution, directed transport can persist even when the net bias vanishes ($\langle \eta(t)\rangle=0$). According to the jump-relaxation framework, the stationary average velocity at zero mean amplitude obeys:
\begin{equation}
    \langle v\rangle\Big|_{\zeta=0}=\lambda\langle\Delta x_R\rangle.
    \label{ratchet_condition}
\end{equation}
A non-vanishing average relaxation distance $\langle \Delta x_R\rangle$ leads to directed transport in the absence of a net force ($\langle\eta(t)\rangle=0$), i.e., a manifestation of the ratchet effect. Consequently, the observed transport anomaly depends on how the average force is rescaled. The requirement
\begin{equation}
    \langle \Delta x_R\rangle\Big|_{\zeta=0}\neq 0
\end{equation}
imposes specific constraints on both the amplitude distribution $\rho(z)$ and the stochastic driving mechanism. First, the active fluctuations must be bidirectional to occur when $\zeta=0$, i.e. both positive and negative fluctuations must be possible. In case of unidirectional amplitude distribution all moments $\langle z_i^n\rangle$ vanish in the limit $\zeta \to0$. Second, by analogy with the deconstruction of absolute negative mobility, the amplitude distribution $\rho(z)$ must be asymmetric to satisfy this inequality, i.e. the probabilities of positive ($P_+$) and negative ($P_-$) relaxations must be unequal as was shown in the right panel of Fig.~\ref{fig:jump}. Systems operating under these conditions are classified as asymmetrically tilting ratchets \cite{reimann2002brownian}.

As established in Chapter~\ref{GiantEnhancement_chapt}, the geometric bounds of the periodic potential $U(x)$ constrain the mean relaxation distance ($\langle \Delta x_R\rangle < L/2$). By contrast, the mean spiking rate $\lambda$ in Eq.~\eqref{ratchet_condition} can vary over a broader parameter range, providing a stronger mechanism to amplify the magnitude of the ratchet velocity, while the sign of $\langle \Delta x_R\rangle$ sets the transport direction. Note, that mean spiking rate $\lambda$ cannot be increased indefinitely as the magnitude of the average relaxation is its monotonically decreasing function $|\langle \Delta x_R\rangle|(\lambda)$, i.e $\langle \Delta x_R\rangle|_{\lambda\to\infty}=0$.

Figure~\ref{fig:ratchet} compares the two scaling implementations of the active force $\eta(t)$ across two values of the variance $\sigma^2$. As the average amplitude $\zeta$ varies, the mean relaxation distance evolves periodically with spatial period $L$, whereas the baseline term $\lambda\zeta$ increases linearly. Consequently, the average velocity $\langle v\rangle$ oscillates around the free-particle baseline $v_0 = \lambda\zeta$. At larger variance ($\sigma^2 = 20$), the ratchet effect persists even for rare fluctuations ($\lambda = 1$) because higher variance enables non-zero barrier-crossing probabilities.

In the rare fluctuation regime ($\lambda \to 0$) the emergence of the ratchet effect is tied to the variance $\sigma^2$ of the amplitude distribution $\rho(z)$ in addition to the two conditions of bidirectionality and asymmetry of fluctuations. When the variance is small enough so that the individual impulses lack sufficient energy to induce potential barrier crossings, i.e. $p_\pm=0$, where we used the definition of these probabilities in Eq.~\eqref{prob_pm}. Because the mean interval between fluctuations is much longer than the intra-well relaxation time ($\tau_P \gg \tau_R$), the particle relaxes to the local potential minimum between successive $\delta$-impulses. In this limit, the relaxation distance $\Delta x_R(z)$ reduces to:
\begin{equation}
    \Delta x_R(z)\Big|_{\lambda\to 0}=-z.
\end{equation}
Because $\langle z\rangle = \zeta = 0$, the mean relaxation distance vanishes:
\begin{equation}
    \langle \Delta x_R\rangle\Big|_{\lambda\to 0}=\langle -z\rangle=-\zeta=0.
\end{equation}
Therefore, at least one barrier-crossing probability ($p_+ \neq 0$ or $p_- \neq 0$) must be non-zero to generate a net ratchet current when fluctuations are rare. This restriction vanishes at higher spiking rates, where multi-impulse dynamics dominate, i.e. another fluctuation may occur during the relaxation towards potential minimum.

Because this model serves as a paradigm in nonequilibrium statistical mechanics \cite{risken1989fokker}, these results apply across a broad spectrum of physical and biological systems, including Josephson junctions, cold atoms in optical lattices, and intracellular motor transport \cite{ezber2020dynein,ariga2021noise}. Direct experimental implementations can be realized using optical tweezer platforms \cite{park2020rapid,paneru2021transport}.

\section{Conclusions}

In this thesis, we have outlined the general characteristics of thermal fluctuations and contrasted them with inherently nonequilibrium active fluctuations. We have established the fundamentals of Brownian particle transport in periodic potentials across both the overdamped and inertial regimes. The introductory portion of this thesis concluded with an overview of the transport anomalies directly relevant to our main results.

The central contribution of this work is the discovery of a novel transport mechanism in periodic systems featuring steep potential landscapes. By harnessing deterministic relaxation toward potential minima, one can robustly control the transport of a Brownian particle simply by tailoring the amplitude distribution of active fluctuations. For a properly tuned, skew-normal amplitude distribution, directed velocity can be enhanced by several orders of magnitude compared to the free-particle limit.

This counterintuitive phenomenon was elucidated both qualitatively and quantitatively using a phenomenological jump-relaxation model. Within this framework, particle displacement is decomposed into two distinct stages: an instantaneous ``jump'' driven by active fluctuations and a subsequent ``relaxation'' down the potential landscape. We presented several levels of complexity for this model, ranging from a coarse-grained variant, which accurately predicts the asymptotic average velocity in high potential barrier limits ($\varepsilon \gg 1$) to a comprehensive framework valid across all parameter regimes.

Because this transport mechanism was initially uncovered in the overdamped limit, it was essential to examine its stability under inertial dynamics. We demonstrated that, in specific parameter regimes, varying particle inertia can directly induce transport enhancement. The jump-relaxation model was successfully extended to inertial dynamics; specifically, the coupling between the amplitude distribution $\rho(z)$ and the spatial period $L$ shifted to a coupling between $\rho(z)$ and the potential energy landscape $U(x)$.

Mapping the global influence of inertia over parameter space revealed that mass can enhance transport by either increasing the magnitude of directed velocity or inducing directed motion in parameter regions where it is absent in the overdamped limit. Conversely, at lower amplitude variance $\sigma^2$, inertia generally suppresses transport enhancement, either dampening or completely extinguishing the effect.

In the final results chapters, the influence of the active fluctuation scaling scheme was thoroughly investigated. To vary the average bias exerted by the fluctuations, one can adjust either the mean spiking rate $\lambda$ or the average amplitude $\zeta$. Under rate scaling, absolute negative mobility (ANM) was uncovered, establishing what is currently the most minimal model known to exhibit this phenomenon, i.e., the inertial Brownian particle under periodic driving in a nonlinear periodic potential was replaced by an overdamped system driven by asymmetric active fluctuations. Under fluctuation amplitude scaling, the classic ratchet effect emerged instead.

The theoretical predictions presented in this work can be tested experimentally using Josephson junctions \cite{nagel2008observation} or cold atoms in optical lattices \cite{lutz2013beyond}. However, a colloidal particle trapped in optical tweezers \cite{park2020rapid,paneru2021transport} offers the most direct experimental platform: by dynamically displacing the optical trap center, one can faithfully recreate arbitrary active force profiles in a controlled environment.

\section{Outline}

Because a Brownian particle in a periodic potential serves as a paradigmatic model in statistical physics, we anticipate that the insights and methodology established in this thesis will inspire follow-up research into active-fluctuation-driven transport in more complex mesoscopic systems. This opens several promising avenues for future investigation.

First, replacing the standard sinusoidal potential with alternative periodic profiles could uncover novel transport phenomena, particularly in the presence of inertial dynamics. Furthermore, the translational symmetry of the system would be broken if a disordered potential were employed. This could significantly impact the transport mechanism identified in this thesis, as it relies fundamentally on the spatial periodicity of the potential landscape. 

While this thesis focused primarily on the stationary average velocity $\langle v\rangle$, exploring the diffusive properties of the system represents a natural extension. Analyzing the time-dependent diffusion coefficient,
\begin{equation}
    D(t) = \frac{\sigma^2_x(t)}{2t},
\end{equation}
could expose distinct anomalies regimes at short and intermediate times corresponding to, e.g., the mean spiking rate $\lambda$ or the relaxation time $\tau_R$. 

Furthermore, evaluating transport efficiency \cite{spiechowicz2014brownian,spiechowicz2015efficiency} by comparing the energy injected by active fluctuations against the mechanical power output would provide a deeper thermodynamic perspective. This could be complemented by investigating fundamental thermodynamic quantities, such as total entropy production and the entropy production rate, yielding a much more comprehensive physical picture of the non-equilibrium steady state.

Finally, the driving mechanism itself can be generalized. Replacing the white Poisson shot noise with alternative models of active fluctuations in particular correlated ones, such as colored Poisson noise, dichotomous noise or kangaroo fluctuations could introduce non-Markovian memory effects and uncover entirely new classes of transport anomalies.

\printbibliography[heading=bibintoc]

@article{brown1828xxvii,
  title={XXVII. A brief account of microscopical observations made in the months of June, July and August 1827, on the particles contained in the pollen of plants; and on the general existence of active molecules in organic and inorganic bodies},
  author={Brown, Robert},
  journal={The Philosophical Magazine},
  volume={4},
  number={21},
  pages={161--173},
  year={1828},
  publisher={Taylor \& Francis}
}

@article{reimann1996brownian,
  title={Brownian motors driven by temperature oscillations},
  author={Reimann, Peter and Bartussek, Roland and H{\"a}ussler, Ralf and H{\"a}nggi, Peter},
  journal={Physics Letters A},
  volume={215},
  number={1-2},
  pages={26--31},
  year={1996},
  publisher={Elsevier}
}

@book{ashcroft1976solid,
  title={Solid State Physics},
  author={Ashcroft, Neil W. and Mermin, N. David},
  year={1976},
  publisher={Saunders College},
  address={Philadelphia}
}

@article{franosch2011resonances,
  title={Resonances arising from hydrodynamic memory in Brownian motion},
  author={Franosch, Thomas and Grimm, Matthias and Belushkin, Maxim and Mor, Flavio M. and Foffi, Giuseppe and Forr{\'o}, L{\'a}szl{\'o} and Jeney, Sylvia},
  journal={Nature},
  volume={478},
  number={7367},
  pages={85--88},
  year={2011},
  publisher={Nature Publishing Group UK London}
}

@article{bartussek1994periodically,
  title={Periodically rocked thermal ratchets},
  author={Bartussek, Roland and H{\"a}nggi, Peter and Kissner, J{\"u}rgen G},
  journal={Europhysics Letters},
  volume={28},
  number={7},
  pages={459--464},
  year={1994}
}

@article{mu2009enhanced,
  title={Enhanced particle transport in an oscillating sinusoidal optical potential},
  author={Mu, Weiqiang and Liu, Zhifu and Luan, Lan and Wang, Guisheng and Spalding, Gabe and Ketterson, John B.},
  journal={New Journal of Physics},
  volume={11},
  number={10},
  pages={103017},
  year={2009}
}

@article{astumian1997thermodynamics,
  title={Thermodynamics and kinetics of a Brownian motor},
  author={Astumian, R. Dean},
  journal={Science},
  volume={276},
  number={5314},
  pages={917--922},
  year={1997},
  publisher={American Association for the Advancement of Science}
}

@article{linke2006self,
  title={Self-propelled Leidenfrost droplets},
  author={Linke, Heiner and Alem{\'a}n, Benjamin J. and Melling, Laura D. and Taormina, Michael J. and Francis, Mathew J. and Dow-Hygelund, Corey C. and Narayanan, Vinod and Taylor, Richard P. and Stout, Andrew},
  journal={Physical Review Letters},
  volume={96},
  number={15},
  pages={154502},
  year={2006},
  publisher={APS}
}

@article{villegas2003superconducting,
  title={A superconducting reversible rectifier that controls the motion of magnetic flux quanta},
  author={Villegas, Javier E. and Savel'ev, Sergey and Nori, Franco and Gonzalez, Elvira M. and Anguita, José V. and Garcia, R and Vicent, JL},
  journal={Science},
  volume={302},
  number={5648},
  pages={1188--1191},
  year={2003},
  publisher={American Association for the Advancement of Science}
}

@article{rousselet1994directional,
  title={Directional motion of Brownian particles induced by a periodic asymmetric potential},
  author={Rousselet, Juliette and Salome, Laurence and Ajdari, Armand and Prostt, Jacques},
  journal={Nature},
  volume={370},
  number={6489},
  pages={446--447},
  year={1994},
  publisher={Nature Publishing Group UK London}
}

@article{faucheux1995optical,
  title={Optical thermal ratchet},
  author={Faucheux, Luc P. and Bourdieu, Laurent and Kaplan, Peter D. and Libchaber, Albert J.},
  journal={Physical Review Letters},
  volume={74},
  number={9},
  pages={1504},
  year={1995},
  publisher={APS}
}

@article{van1999brownian,
  title={Brownian ratchets: Molecular separations in lipid bilayers supported on patterned arrays},
  author={Van Oudenaarden, Alexander and Boxer, Steven G.},
  journal={Science},
  volume={285},
  number={5430},
  pages={1046--1048},
  year={1999},
  publisher={American Association for the Advancement of Science}
}

@article{mcfaul2012cell,
  title={Cell separation based on size and deformability using microfluidic funnel ratchets},
  author={McFaul, Sarah M and Lin, Bill K. and Ma, Hongshen},
  journal={Lab on a Chip},
  volume={12},
  number={13},
  pages={2369--2376},
  year={2012},
  publisher={The Royal Society of Chemistry}
}

@article{hu2017brownian,
  title={Brownian ratchet mechanism for faithful segregation of low-copy-number plasmids},
  author={Hu, Longhua and Vecchiarelli, Anthony G. and Mizuuchi, Kiyoshi and Neuman, Keir C. and Liu, Jian},
  journal={Biophysical Journal},
  volume={112},
  number={7},
  pages={1489--1502},
  year={2017},
  publisher={Elsevier}
}

@article{rusch2026intermediate,
  title={Intermediate scattering function of Brownian particles in a tilted cosine potential},
  author={Rusch, Regina and Franosch, Thomas},
  journal={Physical Review E},
  volume={114},
  number={1},
  pages={014127},
  year={2026},
  publisher={APS}
}

@article{luczka2005non,
  title={Non-Markovian stochastic processes: Colored noise},
  author={{\L}uczka, Jerzy},
  journal={Chaos: An Interdisciplinary Journal of Nonlinear Science},
  volume={15},
  number={2},
  year={2005},
  publisher={AIP Publishing}
}

@article{mondal2025role,
  title={Role of noise-modulated self-propulsion in driving spatiotemporal orders in active systems},
  author={Mondal, Kaustav and Maiti, Tarpan and Ghosh, Pushpita},
  journal={Journal of Chemical Theory and Computation},
  volume={21},
  number={9},
  pages={4405--4416},
  year={2025},
  publisher={ACS Publications}
}

@article{romanczuk2012active,
  title={Active Brownian particles: From individual to collective stochastic dynamics},
  author={Romanczuk, Pawel and B{\"a}r, Markus and Ebeling, Werner and Lindner, Benjamin and Schimansky-Geier, Lutz},
  journal={The European Physical Journal Special Topics},
  volume={202},
  number={1},
  pages={1--162},
  year={2012},
  publisher={Springer}
}

@article{romanczuk2010quasideterministic,
  title={Quasideterministic transport of Brownian particles in an oscillating periodic potential},
  author={Romanczuk, Pawel and M{\"u}ller, Felix and Schimansky-Geier, Lutz},
  journal={Physical Review E},
  volume={81},
  number={6},
  pages={061120},
  year={2010},
  publisher={APS}
}

@article{einstein1905molekularkinetischen,
  title={{\"U}ber die von der molekularkinetischen Theorie der W{\"a}rme geforderte Bewegung von in ruhenden Fl{\"u}ssigkeiten suspendierten Teilchen},
  author={Einstein, Albert},
  journal={Annalen der Physik},
  volume={4},
  year={1905}
}

@book{risken1989fokker,
  title={The Fokker-Planck Equation: Methods of Solution and Applications},
  author={Risken, H. and Frank, T.},
  isbn={9783540615309},
  lccn={96033182},
  series={Springer Series in Synergetics},
  %url={https://books.google.pl/books?id=MG2V9vTgSgEC},
  year={1996},
  publisher={Springer Berlin Heidelberg}
}

@article{von1906kinetischen,
  title={Zur kinetischen Theorie der Brownschen Molekularbewegung und der Suspensionen},
  author={von Smoluchowski, Marian},
  journal={Annalen der Physik},
  volume={326},
  number={14},
  pages={756--780},
  year={1906},
  publisher={Wiley Online Library}
}

@article{perrin1909mouvement,
  title={Mouvement brownien et r{\'e}alit{\'e} mol{\'e}culaire},
  author={Perrin, Jean},
  journal={Annales de Chimie et de Physique},
  year={1909},
  publisher={Masson et Cie}
}

@article{schottky1918spontane,
  title={{\"U}ber spontane Stromschwankungen in verschiedenen Elektrizit{\"a}tsleitern},
  author={Schottky, Walter},
  journal={Annalen der Physik},
  volume={362},
  number={23},
  pages={541--567},
  year={1918},
  publisher={WILEY-VCH Verlag Leipzig}
}

@article{johnson1928thermal,
  title={Thermal agitation of electricity in conductors},
  author={Johnson, John Bertrand},
  journal={Physical Review},
  volume={32},
  number={1},
  pages={97--109},
  year={1928}
}

@article{nyquist1928thermal,
  title={Thermal agitation of electric charge in conductors},
  author={Nyquist, Harry},
  journal={Physical Review},
  volume={32},
  number={1},
  pages={110},
  year={1928},
  publisher={APS}
}

@article{callen1951irreversibility,
  title={Irreversibility and generalized noise},
  author={Callen, Herbert B. and Welton, Theodore A.},
  journal={Physical Review},
  volume={83},
  number={1},
  pages={34},
  year={1951},
  publisher={APS}
}

@article{ashkin1970acceleration,
  title={Acceleration and trapping of particles by radiation pressure},
  author={Ashkin, Arthur},
  journal={Physical Review Letters},
  volume={24},
  number={4},
  pages={156},
  year={1970},
  publisher={APS}
}

@article{ashkin1986observation,
  title={Observation of a single-beam gradient force optical trap for dielectric particles},
  author={Ashkin, Arthur and Dziedzic, James M. and Bjorkholm, John E and Chu, Steven},
  journal={Optics Letters},
  volume={11},
  number={5},
  pages={288--290},
  year={1986},
  publisher={Optical Society of America}
}

@article{neuman2004optical,
  title={Optical trapping},
  author={Neuman, Keir C. and Block, Steven M.},
  journal={Review of Scientific Instruments},
  volume={75},
  number={9},
  pages={2787--2809},
  year={2004},
  publisher={American Institute of Physics}
}

@article{norregaard2017manipulation,
  title={Manipulation and motion of organelles and single molecules in living cells},
  author={Norregaard, Kamilla and Metzler, Ralf and Ritter, Christine M. and Berg-S{\o}rensen, Kirstine and Oddershede, Lene B.},
  journal={Chemical Reviews},
  volume={117},
  number={5},
  pages={4342--4375},
  year={2017},
  publisher={ACS Publications}
}

@article{arbore2019probing,
  title={Probing force in living cells with optical tweezers: from single-molecule mechanics to cell mechanotransduction},
  author={Arbore, Claudia and Perego, Laura and Sergides, Marios and Capitanio, Marco},
  journal={Biophysical Reviews},
  volume={11},
  number={5},
  pages={765--782},
  year={2019},
  publisher={Springer}
}

@article{diekmann2016nanoscopy,
  title={Nanoscopy of bacterial cells immobilized by holographic optical tweezers},
  author={Diekmann, Robin and Wolfson, Deanna L. and Spahn, Christoph and Heilemann, Mike and Sch{\"u}ttpelz, Mark and Huser, Thomas},
  journal={Nature Communications},
  volume={7},
  number={1},
  pages={13711},
  year={2016},
  publisher={Nature Publishing Group UK London}
}

@article{rohrbach2002trapping,
  title={Trapping forces, force constants, and potential depths for dielectric spheres in the presence of spherical aberrations},
  author={Rohrbach, Alexander and Stelzer, Ernst H.K.},
  journal={Applied Optics},
  volume={41},
  number={13},
  pages={2494--2507},
  year={2002},
  publisher={Optical Society of America}
}

@article{juan2011plasmon,
  title={Plasmon nano-optical tweezers},
  author={Juan, Mathieu L. and Righini, Maurizio and Quidant, Romain},
  journal={Nature Photonics},
  volume={5},
  number={6},
  pages={349--356},
  year={2011},
  publisher={Nature Publishing Group UK London}
}

@article{li2010measurement,
  title={Measurement of the instantaneous velocity of a Brownian particle},
  author={Li, Tongcang and Kheifets, Simon and Medellin, David and Raizen, Mark G.},
  journal={Science},
  volume={328},
  number={5986},
  pages={1673--1675},
  year={2010},
  publisher={American Association for the Advancement of Science}
}

@article{huang2011direct,
  title={Direct observation of the full transition from ballistic to diffusive Brownian motion in a liquid},
  author={Huang, Rongxin and Chavez, Isaac and Taute, Katja M. and Luki{\'c}, Branimir and Jeney, Sylvia and Raizen, Mark G. and Florin, Ernst-Ludwig},
  journal={Nature Physics},
  volume={7},
  number={7},
  pages={576--580},
  year={2011},
  publisher={Nature Publishing Group UK London}
}

@article{kumar2018nanoscale,
  title={Nanoscale virtual potentials using optical tweezers},
  author={Kumar, Avinash and Bechhoefer, John},
  journal={Applied Physics Letters},
  volume={113},
  number={18},
  year={2018},
  publisher={AIP Publishing}
}

@article{albay2018optical,
  title={Optical tweezers as a mathematically driven spatio-temporal potential generator},
  author={Albay, John AC and Paneru, Govind and Pak, Hyuk Kyu and Jun, Yonggun},
  journal={Optics Express},
  volume={26},
  number={23},
  pages={29906--29915},
  year={2018},
  publisher={Optical Society of America}
}

@article{paneru2023bona,
  title={Bona fide stochastic resonance under nonGaussian active fluctuations},
  author={Paneru, Govind and Tlusty, Tsvi and Pak, Hyuk Kyu},
  journal={Soft Matter},
  volume={19},
  number={7},
  pages={1356--1362},
  year={2023},
  publisher={The Royal Society of Chemistry}
}

@book{sekimoto2010stochastic,
  title={Stochastic Energetics},
  author={Sekimoto, Ken},
  isbn={9783642054112},
  lccn={2009943129},
  series={Lecture Notes in Physics},
  %url={https://books.google.pl/books?id=8Fq7BQAAQBAJ},
  year={2010},
  publisher={Springer Berlin Heidelberg}
}

@article{vilela2024dynamics,
  title={Dynamics and sorting of run-and-tumble particles in fluid flows with transport barriers},
  author={Vilela, Rafael Dias and Grados, Alfredo J. and Angilella, Jean-R{\'e}gis},
  journal={Journal of Physics: Complexity},
  volume={5},
  number={3},
  pages={035003},
  year={2024},
  publisher={IOP Publishing}
}

@article{lee2022effects,
  title={Effects of the non-Markovianity and non-Gaussianity of active environmental noises on engine performance},
  author={Lee, Jae Sung and Park, Hyunggyu},
  journal={Physical Review E},
  volume={105},
  number={2},
  pages={024130},
  year={2022},
  publisher={APS}
}

@article{park2020rapid,
  title={Rapid-prototyping a Brownian particle in an active bath},
  author={Park, Jin Tae and Paneru, Govind and Kwon, Chulan and Granick, Steve and Pak, Hyuk Kyu},
  journal={Soft Matter},
  volume={16},
  number={35},
  pages={8122--8127},
  year={2020},
  publisher={Royal Society of Chemistry}
}

@article{ariga2021noise,
  title={Noise-induced acceleration of single molecule kinesin-1},
  author={Ariga, Takayuki and Tateishi, Keito and Tomishige, Michio and Mizuno, Daisuke},
  journal={Physical Review Letters},
  volume={127},
  number={17},
  pages={178101},
  year={2021},
  publisher={APS}
}

@article{lutz2013beyond,
  title={Beyond Boltzmann--Gibbs statistical mechanics in optical lattices},
  author={Lutz, Eric and Renzoni, Ferruccio},
  journal={Nature Physics},
  volume={9},
  number={10},
  pages={615--619},
  year={2013},
  publisher={Nature Publishing Group UK London}
}

@article{paneru2021transport,
  title={Transport and diffusion enhancement in experimentally realized non-Gaussian correlated ratchets},
  author={Paneru, Govind and Park, Jin Tae and Pak, Hyuk Kyu},
  journal={The Journal of Physical Chemistry Letters},
  volume={12},
  number={45},
  pages={11078--11084},
  year={2021},
  publisher={ACS Publications}
}

@article{ezber2020dynein,
  title={Dynein harnesses active fluctuations of microtubules for faster movement},
  author={Ezber, Yasin and Belyy, Vladislav and Can, Sinan and Yildiz, Ahmet},
  journal={Nature Physics},
  volume={16},
  number={3},
  pages={312--316},
  year={2020},
  publisher={Nature Publishing Group UK London}
}

@article{gutierrez2018induced,
  title={Induced clustering of Escherichia coli by acoustic fields},
  author={Guti{\'e}rrez-Ramos, Salom{\'e} and Hoyos, Mauricio and Ruiz-Su{\'a}rez, J.C.},
  journal={Scientific Reports},
  volume={8},
  number={1},
  pages={4668},
  year={2018},
  publisher={Nature Publishing Group UK London}
}

@article{tung2017fluid,
  title={Fluid viscoelasticity promotes collective swimming of sperm},
  author={Tung, Chih-kuan and Lin, Chungwei and Harvey, Benedict and Fiore, Alyssa G. and Ardon, Florencia and Wu, Mingming and Suarez, Susan S.},
  journal={Scientific Reports},
  volume={7},
  number={1},
  pages={3152},
  year={2017},
  publisher={Nature Publishing Group UK London}
}

@article{liebchen2022chiral,
  title={Chiral active matter},
  author={Liebchen, Benno and Levis, Demian},
  journal={Europhysics Letters},
  volume={139},
  number={6},
  pages={67001},
  year={2022},
  publisher={EDP Sciences, IOP Publishing and Societ{\`a} Italiana di Fisica}
}

@book{gardiner1985handbook,
  title={Handbook of Stochastic Methods for Physics, Chemistry and the Natural Sciences},
  author={Gardiner, Crispin W.},
  journal={Springer series in synergetics},
  year={1985},
  publisher={Springer Berlin Heidelberg}
}

@article{das2015boundaries,
  title={Boundaries can steer active Janus spheres},
  author={Das, Sambeeta and Garg, Astha and Campbell, Andrew I. and Howse, Jonathan and Sen, Ayusman and Velegol, Darrell and Golestanian, Ramin and Ebbens, Stephen J.},
  journal={Nature Communications},
  volume={6},
  number={1},
  pages={8999},
  year={2015},
  publisher={Nature Publishing Group UK London}
}

@article{dabelow2019irreversibility,
  title={Irreversibility in active matter systems: Fluctuation theorem and mutual information},
  author={Dabelow, Lennart and Bo, Stefano and Eichhorn, Ralf},
  journal={Physical Review X},
  volume={9},
  number={2},
  pages={021009},
  year={2019},
  publisher={APS}
}

@article{bier2003processive,
  title={Processive motor protein as an overdamped Brownian stepper},
  author={Bier, Martin},
  journal={Physical Review Letters},
  volume={91},
  number={14},
  pages={148104},
  year={2003},
  publisher={APS}
}

@article{spiechowicz2015gpu,
  title={GPU accelerated Monte Carlo simulation of Brownian motors dynamics with CUDA},
  author={Spiechowicz, Jakub and Kostur, Marcin and Machura, Łukasz},
  journal={Computer Physics Communications},
  volume={191},
  pages={140--149},
  year={2015},
  publisher={Elsevier}
}

@article{o2022time,
  title={Time irreversibility in active matter, from micro to macro},
  author={O’Byrne, J{\'e}r{\'e}my and Kafri, Yariv and Tailleur, Julien and van Wijland, Fr{\'e}d{\'e}ric},
  journal={Nature Reviews Physics},
  volume={4},
  number={3},
  pages={167--183},
  year={2022},
  publisher={Nature Publishing Group UK London}
}

@article{maggi2017memory,
  title={Memory-less response and violation of the fluctuation-dissipation theorem in colloids suspended in an active bath},
  author={Maggi, Claudio and Paoluzzi, Matteo and Angelani, Luca and Di Leonardo, Roberto},
  journal={Scientific Reports},
  volume={7},
  number={1},
  pages={17588},
  year={2017},
  publisher={Nature Publishing Group UK London}
}

@article{vutukuri2016dynamic,
  title={Dynamic self-organization of side-propelling colloidal rods: experiments and simulations},
  author={Vutukuri, Hanumantha Rao and Preisler, Zden{\v{e}}k and Besseling, Thijs H. and Van Blaaderen, Alfons and Dijkstra, Marjolein and Huck, Wilhelm T.S.},
  journal={Soft Matter},
  volume={12},
  number={48},
  pages={9657--9665},
  year={2016},
  publisher={Royal Society of Chemistry}
}

@article{gilbert1995pathway,
  title={Pathway of processive ATP hydrolysis by kinesin},
  author={Gilbert, Susan P. and Webb, Martin R. and Brune, Martin and Johnson, Kenneth A.},
  journal={Nature},
  volume={373},
  number={6516},
  pages={671--676},
  year={1995},
  publisher={Nature Publishing Group UK London}
}

@article{ross2006processive,
  title={Processive bidirectional motion of dynein--dynactin complexes in vitro},
  author={Ross, Jennifer L. and Wallace, Karen and Shuman, Henry and Goldman, Yale E. and Holzbaur, Erika L.F.},
  journal={Nature Cell Biology},
  volume={8},
  number={6},
  pages={562--570},
  year={2006},
  publisher={Nature Publishing Group UK London}
}

@article{golestanian2005propulsion,
  title={Propulsion of a molecular machine by asymmetric distribution of reaction products},
  author={Golestanian, Ramin and Liverpool, Tanniemola B. and Ajdari, Armand},
  journal={Physical Review Letters},
  volume={94},
  number={22},
  pages={220801},
  year={2005},
  publisher={APS}
}

@article{ye2020active,
  title={Active noise experienced by a passive particle trapped in an active bath},
  author={Ye, Simin and Liu, Peng and Ye, Fangfu and Chen, Ke and Yang, Mingcheng},
  journal={Soft Matter},
  volume={16},
  number={19},
  pages={4655--4660},
  year={2020},
  publisher={Royal Society of Chemistry}
}

@article{viterbi1963phase,
  title={Phase-locked loop dynamics in the presence of noise by Fokker-Planck techniques},
  author={Viterbi, Andrew J.},
  journal={Proceedings of the IEEE},
  volume={51},
  number={12},
  pages={1737--1753},
  year={1963},
  publisher={IEEE}
}

@article{bellando2022giant,
  title={Giant diffusion of nanomechanical rotors in a tilted washboard potential},
  author={Bellando, Louis and Kleine, Melissa and Amarouchene, Yacine and Perrin, Mathias and Louyer, Yann},
  journal={Physical Review Letters},
  volume={129},
  number={2},
  pages={023602},
  year={2022},
  publisher={APS}
}

@article{plan2020active,
  title={Active matter in a viscoelastic environment},
  author={Plan, Emmanuel L.C. V.I. M. and Yeomans, Julia M. and Doostmohammadi, Amin},
  journal={Physical Review Fluids},
  volume={5},
  number={2},
  pages={023102},
  year={2020},
  publisher={APS}
}

@article{hwang2017quantifying,
  title={Quantifying the heat dissipation from a molecular motor’s transport properties in nonequilibrium steady states},
  author={Hwang, Wonseok and Hyeon, Changbong},
  journal={The Journal of Physical Chemistry Letters},
  volume={8},
  number={1},
  pages={250--256},
  year={2017},
  publisher={ACS Publications}
}

@article{seyforth2022nonequilibrium,
  title={Nonequilibrium fluctuations and nonlinear response of an active bath},
  author={Seyforth, Hunter and Gomez, Mauricio and Rogers, W. Benjamin and Ross, Jennifer L. and Ahmed, Wylie W.},
  journal={Physical Review Research},
  volume={4},
  number={2},
  pages={023043},
  year={2022},
  publisher={APS}
}

@article{spiechowicz2015efficiency,
  title={Efficiency of the SQUID ratchet driven by external current},
  author={Spiechowicz, Jakub and {\L}uczka, Jerzy},
  journal={New Journal of Physics},
  volume={17},
  number={2},
  pages={023054},
  year={2015},
  publisher={IOP Publishing}
}

@article{howse2007self,
  title={Self-motile colloidal particles: from directed propulsion to random walk},
  author={Howse, Jonathan R. and Jones, Richard A.L. and Ryan, Anthony J. and Gough, Tim and Vafabakhsh, Reza and Golestanian, Ramin},
  journal={Physical Review Letters},
  volume={99},
  number={4},
  pages={048102},
  year={2007},
  publisher={APS}
}

@article{prost2015active,
  title={Active gel physics},
  author={Prost, Jacques and J{\"u}licher, Frank and Joanny, Jean-Fran{\c{c}}ois},
  journal={Nature Physics},
  volume={11},
  number={2},
  pages={111--117},
  year={2015},
  publisher={Nature Publishing Group UK London}
}

@article{wang2012nano,
  title={Nano/microscale motors: biomedical opportunities and challenges},
  author={Wang, Joseph and Gao, Wei},
  journal={ACS Nano},
  volume={6},
  number={7},
  pages={5745--5751},
  year={2012},
  publisher={ACS Publications}
}

@article{spiechowicz2014brownian,
  title={Brownian motors in the microscale domain: Enhancement of efficiency by noise},
  author={Spiechowicz, Jakub and H{\"a}nggi, Peter and {\L}uczka, Jerzy},
  journal={Physical Review E},
  volume={90},
  number={3},
  pages={032104},
  year={2014},
  publisher={APS}
}

@article{marconi2008fluctuation,
  title={Fluctuation--dissipation: response theory in statistical physics},
  author={Marconi, Umberto M.B. and Puglisi, Andrea and Rondoni, Lamberto and Vulpiani, Angelo},
  journal={Physics Reports},
  volume={461},
  number={4-6},
  pages={111--195},
  year={2008},
  publisher={Elsevier}
}

@article{fodor2016far,
  title={How far from equilibrium is active matter?},
  author={Fodor, {\'E}tienne and Nardini, Cesare and Cates, Michael E. and Tailleur, Julien and Visco, Paolo and Van Wijland, Fr{\'e}d{\'e}ric},
  journal={Physical Review Letters},
  volume={117},
  number={3},
  pages={038103},
  year={2016},
  publisher={APS}
}

@article{speer2007transient,
  title={Transient chaos induces anomalous transport properties of an underdamped Brownian particle},
  author={Speer, David and Eichhorn, Ralf and Reimann, Peter},
  journal={Physical Review E},
  volume={76},
  number={5},
  pages={051110},
  year={2007},
  publisher={APS}
}

@article{slapik2020tunable,
  title={Tunable particle separation via deterministic absolute negative mobility},
  author={S{\l}apik, Aleksandra and Spiechowicz, Jakub},
  journal={Scientific Reports},
  volume={10},
  number={1},
  pages={16639},
  year={2020},
  publisher={Nature Publishing Group UK London}
}

@article{sarracino2016nonlinear,
  title={Nonlinear response of inertial tracers in steady laminar flows: differential and absolute negative mobility},
  author={Sarracino, Alessandro and Cecconi, Fabio and Puglisi, Andrea and Vulpiani, Angelo},
  journal={Physical Review Letters},
  volume={117},
  number={17},
  pages={174501},
  year={2016},
  publisher={APS}
}

@article{machura2007absolute,
  title={Absolute negative mobility induced by thermal equilibrium fluctuations},
  author={Machura, Łukasz and Kostur, Marcin and Talkner, Peter and {\L}uczka, Jerzy and H{\"a}nggi, Peter},
  journal={Physical Review Letters},
  volume={98},
  number={4},
  pages={040601},
  year={2007},
  publisher={APS}
}

@article{rizkallah2023absolute,
  title={Absolute negative mobility of an active tracer in a crowded environment},
  author={Rizkallah, Pierre and Sarracino, Alessandro and B{\'e}nichou, Olivier and Illien, Pierre},
  journal={Physical Review Letters},
  volume={130},
  number={21},
  pages={218201},
  year={2023},
  publisher={APS}
}

@article{zia2002getting,
  title={Getting more from pushing less: Negative specific heat and conductivity in nonequilibrium steady states},
  author={Zia, Royce K.P. and Praestgaard, Eigil Luxh{\o}j and Mouritsen, Ole G.},
  journal={American Journal of Physics},
  volume={70},
  number={4},
  pages={384--392},
  year={2002},
  publisher={American Association of Physics Teachers}
}

@article{ghosh2014giant,
  title={Giant negative mobility of Janus particles in a corrugated channel},
  author={Ghosh, Pulak K. and H{\"a}nggi, Peter and Marchesoni, Fabio and Nori, Franco},
  journal={Physical Review E},
  volume={89},
  number={6},
  pages={062115},
  year={2014},
  publisher={APS}
}

@article{duran1993arching,
  title={Arching effect model for particle size segregation},
  author={Duran, Jacques and Rajchenbach, Jean and Cl{\'e}ment, Eric},
  journal={Physical Review Letters},
  volume={70},
  number={16},
  pages={2431},
  year={1993},
  publisher={APS}
}

@article{nagel2008observation,
  title={Observation of negative absolute resistance in a Josephson junction},
  author={Nagel, Joachim and Speer, David and Gaber, Tobias and Sterck, Andreas and Eichhorn, Ralf and Reimann, Peter and Ilin, Konstantin and Siegel, Michael and Koelle, Dieter and Kleiner, Reinhold},
  journal={Physical Review Letters},
  volume={100},
  number={21},
  pages={217001},
  year={2008},
  publisher={APS}
}

@article{keay1995dynamic,
  title={Dynamic localization, absolute negative conductance, and stimulated, multiphoton emission in sequential resonant tunneling semiconductor superlattices},
  author={Keay, Brian J. and Zeuner, Stefan and Allen Jr, S.J. and Maranowski, Kevin D. and Gossard, Art C. and Bhattacharya, Uddalak and Rodwell, Marc J.W.},
  journal={Physical Review Letters},
  volume={75},
  number={22},
  pages={4102},
  year={1995},
  publisher={APS}
}

@article{eichhorn2005moving,
  title={Moving backward noisily},
  author={Eichhorn, Ralf and Reimann, Peter and Cleuren, Bart and Van den Broeck, Christian},
  journal={Chaos: An Interdisciplinary Journal of Nonlinear Science},
  volume={15},
  number={2},
  year={2005},
  publisher={AIP Publishing}
}

@article{ros2005absolute,
  title={Absolute negative particle mobility},
  author={Ros, Alexandra and Eichhorn, Ralf and Regtmeier, Jan and Duong, Thanh Tu and Reimann, Peter and Anselmetti, Dario},
  journal={Nature},
  volume={436},
  number={7053},
  pages={928--928},
  year={2005},
  publisher={Nature Publishing Group UK London}
}

@article{RevModPhys.85.1143,
title={Hydrodynamics of soft active matter},
  author={Marchetti, M. Cristina and Joanny, Jean-Fran{\c{c}}ois and Ramaswamy, Sriram and Liverpool, Tanniemola B. and Prost, Jacques and Rao, Madan and Simha, R. Aditi},
  journal={Reviews of Modern Physics},
  volume={85},
  number={3},
  pages={1143--1189},
  year={2013},
  publisher={APS}
}

@article{bechinger2016active,
  title={Active particles in complex and crowded environments},
  author={Bechinger, Clemens and Di Leonardo, Roberto and L{\"o}wen, Hartmut and Reichhardt, Charles and Volpe, Giorgio and Volpe, Giovanni},
  journal={Reviews of Modern Physics},
  volume={88},
  number={4},
  pages={045006},
  year={2016},
  publisher={APS}
}

@article{van1982diffusion,
  title={The diffusion approximation for Markov processes},
  author={Van Kampen, Nicolaas Godfried},
  journal={Thermodynamics and Kinetics of Biological Processes},
  pages={181--195},
  year={1982},
  publisher={Walter de Gruyter and Co. Berlin, Germany}
}

@article{kubo1966fluctuation,
  title={The fluctuation-dissipation theorem},
  author={Kubo, Ryogo},
  journal={Reports on Progress in Physics},
  volume={29},
  number={1},
  pages={255--284},
  year={1966}
}

@article{seifert2005entropy,
  title={Entropy production along a stochastic trajectory and an integral fluctuation theorem},
  author={Seifert, Udo},
  journal={Physical Review Letters},
  volume={95},
  number={4},
  pages={040602},
  year={2005},
  publisher={APS}
}

@article{seifert2012stochastic,
  title={Stochastic thermodynamics, fluctuation theorems and molecular machines},
  author={Seifert, Udo},
  journal={Reports on Progress in Physics},
  volume={75},
  number={12},
  pages={126001},
  year={2012},
  publisher={IOP Publishing}
}

@book{van1983stochastic,
  title={Stochastic Processes in Physics and Chemistry},
  author={Van Kampen, Nicolaas G. and Reinhardt, William P.},
  year={1983},
  publisher={American Institute of Physics}
}

@article{alkemade1963non,
  title={Non-linear Brownian movement of a generalized Rayleigh model},
  author={Alkemade, Cornelis T.J. and Van Kampen, Nicolaas G. and Macdonald, David K. Chalmers},
  journal={Proceedings of the Royal Society of London. Series A. Mathematical and Physical Sciences},
  volume={271},
  number={1347},
  pages={449--471},
  year={1963},
  publisher={The Royal Society London}
}

@book{feller1991introduction,
  title={An Introduction to Probability Theory and its Applications},
  author={Feller, William},
  volume={2},
  year={1991},
  publisher={John Wiley \& Sons}
}

@article{mabillard2023heat,
  title={Heat fluctuations in chemically active systems},
  author={Mabillard, Jo{\"e}l and Weber, Christoph A and J{\"u}licher, Frank},
  journal={Physical Review E},
  volume={107},
  number={1},
  pages={014118},
  year={2023},
  publisher={APS}
}

@article{hanggi1978derivations,
  title={On derivations and solutions of master equations and asymptotic representations},
  author={H{\"a}nggi, Peter},
  journal={Zeitschrift f{\"u}r Physik B Condensed Matter},
  volume={30},
  number={1},
  pages={85--95},
  year={1978},
  publisher={Springer}
}

@book{huang2009introduction,
  title={Introduction to Statistical Physics, Second Edition},
  author={Huang, Kerson},
  isbn={9780415683678},
  series={Civil and Environmental Engineering},
  year={2009},
  publisher={CRC Press}
}

@article{gnesotto2018broken,
  title={Broken detailed balance and non-equilibrium dynamics in living systems: a review},
  author={Gnesotto, Federico S. and Mura, Federica and Gladrow, Jannes and Broedersz, Chase P.},
  journal={Reports on Progress in Physics},
  volume={81},
  number={6},
  pages={066601},
  year={2018},
  publisher={IOP Publishing}
}

@article{von1912experimental,
  title={Experimentell nachweisbare, der {\"u}blichen Thermodynamik widersprechende Molekularph{\"a}nomene},
  author={von Smoluchowski, Marian},
  journal={Physikalische Zeitschrift},
  volume={13},
  pages={1069},
  year={1912}
}

@book{kanazawa2017statistical,
  title={Statistical mechanics for athermal fluctuation: Non-Gaussian noise in physics},
  author={Kanazawa, Kiyoshi},
  year={2017},
  publisher={Springer}
}

@article{ben2011effective,
  title={Effective temperature of red-blood-cell membrane fluctuations},
  author={Ben-Isaac, Eyal and Park, YongKeun and Popescu, Gabriel and Brown, Frank L.H. and Gov, Nir S and Shokef, Yair},
  journal={Physical Review Letters},
  volume={106},
  number={23},
  pages={238103},
  year={2011},
  publisher={APS}
}

@article{blanter2000shot,
  title={Shot noise in mesoscopic conductors},
  author={Blanter, Ya M. and B{\"u}ttiker, Markus},
  journal={Physics Reports},
  volume={336},
  number={1-2},
  pages={1--166},
  year={2000},
  publisher={Elsevier}
}

@article{behringer2019physics,
  title={The physics of jamming for granular materials: a review},
  author={Behringer, Robert P. and Chakraborty, Bulbul},
  journal={Reports on Progress in Physics},
  volume={82},
  number={1},
  pages={012601},
  year={2019},
  publisher={IOP Publishing}
}

@article{huang2026entropy,
  title={Entropy production in non-Gaussian active matter: A unified fluctuation theorem and deep learning framework},
  author={Huang, Yuanfei and Liu, Chengyu and Miao, Bing and Zhou, Xiang},
  journal={Physical Review Letters},
  volume={136},
  number={6},
  pages={068302},
  year={2026},
  publisher={APS}
}

@article{hanggi2009artificial,
  title={Artificial Brownian motors: Controlling transport on the nanoscale},
  author={H{\"a}nggi, Peter and Marchesoni, Fabio},
  journal={Reviews of Modern Physics},
  volume={81},
  number={1},
  pages={387--442},
  year={2009},
  publisher={APS}
}

@book{gitterman2008noisy,
  title={The Noisy Pendulum},
  author={Gitterman, Moshe},
  year={2008},
  publisher={World Scientific}
}

@article{fulde1975problem,
  title={Problem of Brownian motion in a periodic potential},
  author={Fulde, Peter and Pietronero, Luciano and Schneider, Walter R. and Str{\"a}ssler, Stephan},
  journal={Physical Review Letters},
  volume={35},
  number={26},
  pages={1776},
  year={1975},
  publisher={APS}
}

@article{falo1999ratchet,
  title={Ratchet potential for fluxons in Josephson-junction arrays},
  author={Falo, Fernando and Martinez, Pedro J. and Mazo, Juan J. and Cilla, Sofia},
  journal={Europhysics Letters},
  volume={45},
  number={6},
  pages={700--706},
  year={1999}
}

@article{braun1998nonlinear,
  title={Nonlinear dynamics of the Frenkel--Kontorova model},
  author={Braun, Oleg M. and Kivshar, Yuri S.},
  journal={Physics Reports},
  volume={306},
  number={1-2},
  pages={1--108},
  year={1998},
  publisher={Elsevier}
}

@article{gruner1981nonlinear,
  title={Nonlinear conductivity and noise due to charge-density-wave depinning in $\mathrm{NbSe}_3$},
  author={Gr{\"u}ner, George and Zawadowski, Alfred and Chaikin, Paul M.},
  journal={Physical Review Letters},
  volume={46},
  number={7},
  pages={511},
  year={1981},
  publisher={APS}
}

@article{reimann2002introduction,
  title={Introduction to the physics of Brownian motors},
  author={Reimann, Peter and H{\"a}nggi, Peter},
  journal={Applied Physics A},
  volume={75},
  number={2},
  pages={169--178},
  year={2002},
  publisher={Springer}
}

@article{ten2011brownian,
  title={Brownian motion of a self-propelled particle},
  author={ten Hagen, Borge and van Teeffelen, Sven and L{\"o}wen, Hartmut},
  journal={Journal of Physics: Condensed Matter},
  volume={23},
  number={19},
  pages={194119},
  year={2011}
}

@article{volpe2014simulation,
  title={Simulation of the active Brownian motion of a microswimmer},
  author={Volpe, Giorgio and Gigan, Sylvain and Volpe, Giovanni},
  journal={American Journal of Physics},
  volume={82},
  number={7},
  pages={659--664},
  year={2014},
  publisher={AIP Publishing}
}

@article{caprini2021inertial,
  title={Inertial self-propelled particles},
  author={Caprini, Lorenzo and Marini B.M., Umberto},
  journal={The Journal of Chemical Physics},
  volume={154},
  number={2},
  year={2021},
  publisher={AIP Publishing}
}

@article{lowen2020inertial,
  title={Inertial effects of self-propelled particles: From active Brownian to active Langevin motion},
  author={L{\"o}wen, Hartmut},
  journal={The Journal of Chemical Physics},
  volume={152},
  number={4},
  year={2020},
  publisher={AIP Publishing}
}

@article{reimann2002brownian,
  title={Brownian motors: noisy transport far from equilibrium},
  author={Reimann, Peter},
  journal={Physics Reports},
  volume={361},
  number={2-4},
  pages={57--265},
  year={2002},
  publisher={Elsevier}
}

@book{feynman2013lectures,
  author = {Feynman, Richard P. and Leighton, Robert B. and Sands, Matthew},
  title = {The Feynman Lectures on Physics},
  editor = {Gottlieb, Michael A. and Pfeiffer, Rudolf},
  year = {2013},
  publisher = {Caltech/Basic Books},
  %url = {https://www.feynmanlectures.caltech.edu/},
    note = {New Millennium Edition}
}

@article{uhlenbeck1930theory,
  title={On the theory of the Brownian motion},
  author={Uhlenbeck, George E. and Ornstein, Leonard S.},
  journal={Physical Review},
  volume={36},
  number={5},
  pages={823},
  year={1930},
  publisher={APS}
}

@article{nagai2015collective,
  title={Collective motion of self-propelled particles with memory},
  author={Nagai, Ken H. and Sumino, Yutaka and Montagne, Raul and Aranson, Igor S and Chat{\'e}, Hugues},
  journal={Physical Review Letters},
  volume={114},
  number={16},
  pages={168001},
  year={2015},
  publisher={APS}
}

@article{martin2021statistical,
  title={Statistical mechanics of active Ornstein-Uhlenbeck particles},
  author={Martin, David and O'Byrne, J{\'e}r{\'e}my and Cates, Michael E. and Fodor, {\'E}tienne and Nardini, Cesare and Tailleur, Julien and Van Wijland, Fr{\'e}d{\'e}ric},
  journal={Physical Review E},
  volume={103},
  number={3},
  pages={032607},
  year={2021},
  publisher={APS}
}

@book{stratonovich1967topics,
  title={Topics in the Theory of Random Noise},
  author={Stratonovich, Rouslan L.},
  volume={2},
  year={1967},
  publisher={CRC Press}
}

@article{curie1894symetrie,
  title={Sur la sym{\'e}trie dans les ph{\'e}nom{\`e}nes physiques, sym{\'e}trie d'un champ {\'e}lectrique et d'un champ magn{\'e}tique},
  author={Curie, Pierre},
  journal={Journal de Physique Th{\'e}orique et Appliqu{\'e}e},
  volume={3},
  number={1},
  pages={393--415},
  year={1894},
  publisher={Soci{\'e}t{\'e} Fran{\c{c}}aise de Physique}
}

@article{shen2005nonequilibrium,
  title={Nonequilibrium statistical mechanical models for cytoskeletal assembly: Towards understanding tensegrity in cells},
  author={Shen, Tongye and Wolynes, Peter G.},
  journal={Physical Review E},
  volume={72},
  number={4},
  pages={041927},
  year={2005},
  publisher={APS}
}

@article{rijal2022exact,
  title={Exact distribution of threshold crossing times for protein concentrations: Implication for biological timekeeping},
  author={Rijal, Krishna and Prasad, Ashok and Singh, Abhyudai and Das, Dibyendu},
  journal={Physical Review Letters},
  volume={128},
  number={4},
  pages={048101},
  year={2022},
  publisher={APS}
}

@article{azz,
  title={A class of distributions which includes the normal ones},
  author={Azzalini, Adelchi},
  journal={Scandinavian Journal of Statistics},
  pages={171--178},
  year={1985},
  publisher={JSTOR}
}

@inproceedings{generacja2,
  title={Generating the skew normal random variable},
  author={Ghorbanzadeh, Dariush and Durand, Philippe and Jaupi, Luan},
  booktitle={World Congress on Engineering 2017},
  pages={113--116},
  year={2017}
}

@article{generacja,
  title={A probabilistic representation of the 'skew-normal' distribution},
  author={Henze, Norbert},
  journal={Scandinavian Journal of Statistics},
  pages={271--275},
  year={1986},
  publisher={JSTOR}
}

@book{platen2010numerical,
  title={Numerical Solution of Stochastic Differential Equations with Jumps in Finance},
  author={Platen, Eckhard and Bruti-Liberati, Nicola},
  volume={64},
  year={2010},
  publisher={Springer Science \& Business Media}
}

@article{kim2007numerical,
  title={Numerical method for solving stochastic differential equations with Poissonian white shot noise},
  author={Kim, Changho and Lee, Eok Kyun and H{\"a}nggi, Peter and Talkner, Peter},
  journal={Physical Review E},
  volume={76},
  number={1},
  pages={011109},
  year={2007},
  publisher={APS}
}

@article{mandal2017entropy,
  title={Entropy production and fluctuation theorems for active matter},
  author={Mandal, Dibyendu and Klymko, Katherine and DeWeese, Michael R.},
  journal={Physical Review Letters},
  volume={119},
  number={25},
  pages={258001},
  year={2017},
  publisher={APS}
}

@book{johnson1994,
  title={Continuous Univariate Distributions},
  author={Johnson, Norman L. and Kotz, Samuel and Balakrishnan, Narayanaswamy},
  volume={1},
  year={1994},
  publisher={John Wiley \& Sons}
}

@article{chowdhury2013stochastic,
  title={Stochastic mechano-chemical kinetics of molecular motors: a multidisciplinary enterprise from a physicist’s perspective},
  author={Chowdhury, Debashish},
  journal={Physics Reports},
  volume={529},
  number={1},
  pages={1--197},
  year={2013},
  publisher={Elsevier}
}

@article{fiasconaro2009tuning,
  title={Tuning active Brownian motion with shot-noise energy pulses},
  author={Fiasconaro, Alessandro and Gudowska-Nowak, Ewa and Ebeling, Werner},
  journal={Journal of Statistical Mechanics: Theory and Experiment},
  volume={2009},
  number={01},
  pages={P01029},
  year={2009}
}

@article{hanggi1980langevin,
  title={Langevin description of Markovian integro-differential master equations},
  author={H{\"a}nggi, Peter},
  journal={Zeitschrift f{\"u}r Physik B Condensed Matter},
  volume={36},
  number={3},
  pages={271--282},
  year={1980},
  publisher={Springer}
}

@article{haunggi1994colored,
  title={Colored noise in dynamical systems},
  author={H{\"a}nggi, Peter and Jung, Peter},
  journal={Advances in Chemical Physics},
  volume={89},
  pages={239--326},
  year={1994},
  publisher={Wiley Online Library}
}

@article{woillez2020nonlocal,
  title={Nonlocal stationary probability distributions and escape rates for an active Ornstein--Uhlenbeck particle},
  author={Woillez, Eric and Kafri, Yariv and Lecomte, Vivien},
  journal={Journal of Statistical Mechanics: Theory and Experiment},
  volume={2020},
  number={6},
  pages={063204},
  year={2020},
  publisher={IOP Publishing and SISSA}
}

@article{nguyen2022active,
  title={Active Ornstein--Uhlenbeck model for self-propelled particles with inertia},
  author={Nguyen, Gia Huy Philipp and Wittmann, Ren{\'e} and L{\"o}wen, Hartmut},
  journal={Journal of Physics: Condensed Matter},
  volume={34},
  number={3},
  pages={035101},
  year={2022},
  publisher={IOP Publishing}
}

@article{needleman2017active,
  title={Active matter at the interface between materials science and cell biology},
  author={Needleman, Daniel and Dogic, Zvonimir},
  journal={Nature Reviews Materials},
  volume={2},
  number={9},
  pages={17048},
  year={2017},
  publisher={Nature Publishing Group}
}

@article{mccumber1968effect,
  title={Effect of ac impedance on dc voltage-current characteristics of superconductor weak-link junctions},
  author={McCumber, Dean E.},
  journal={Journal of Applied Physics},
  volume={39},
  number={7},
  pages={3113--3118},
  year={1968},
  publisher={American Institute of Physics}
}

@article{spiechowicz2016transient,
  title={Transient anomalous diffusion in periodic systems: ergodicity, symmetry breaking and velocity relaxation},
  author={Spiechowicz, Jakub and {\L}uczka, Jerzy and H{\"a}nggi, Peter},
  journal={Scientific Reports},
  volume={6},
  number={1},
  pages={30948},
  year={2016},
  publisher={Nature Publishing Group UK London}
}

@article{spiechowicz2022diffusion,
  title={Diffusion coefficient of a Brownian particle in equilibrium and nonequilibrium: Einstein model and beyond},
  author={Spiechowicz, Jakub and Marchenko, Ivan G. and H{\"a}nggi, Peter and {\L}uczka, Jerzy},
  journal={Entropy},
  volume={25},
  number={1},
  pages={42},
  year={2022},
  publisher={MDPI}
}

@article{slapik2019tunable,
  title={Tunable mass separation via negative mobility},
  author={S{\l}apik, Aleksandra and {\L}uczka, Jerzy and H{\"a}nggi, Peter and Spiechowicz, Jakub},
  journal={Physical Review Letters},
  volume={122},
  number={7},
  pages={070602},
  year={2019},
  publisher={APS}
}

@article{slapik2019temperature,
  title={Temperature-induced tunable particle separation},
  author={S{\l}apik, Aleksandra and {\L}uczka, Jerzy and Spiechowicz, Jakub},
  journal={Physical Review Applied},
  volume={12},
  number={5},
  pages={054002},
  year={2019},
  publisher={APS}
}

@article{spiechowicz2019coexistence,
  title={Coexistence of absolute negative mobility and anomalous diffusion},
  author={Spiechowicz, Jakub and H{\"a}nggi, Peter and {\L}uczka, Jerzy},
  journal={New Journal of Physics},
  volume={21},
  number={8},
  pages={083029},
  year={2019},
  publisher={IOP Publishing}
}

@article{spiechowicz2021conundrum,
  title={Conundrum of weak-noise limit for diffusion in a tilted periodic potential},
  author={Spiechowicz, Jakub and {\L}uczka, Jerzy},
  journal={Physical Review E},
  volume={104},
  number={3},
  pages={034104},
  year={2021},
  publisher={APS}
}

@article{spiechowicz2021arcsine,
  title={Arcsine law and multistable Brownian dynamics in a tilted periodic potential},
  author={Spiechowicz, Jakub and {\L}uczka, Jerzy},
  journal={Physical Review E},
  volume={104},
  number={2},
  pages={024132},
  year={2021},
  publisher={APS}
}

@article{spiechowicz2022velocity,
  title={Velocity multistability vs. ergodicity breaking in a biased periodic potential},
  author={Spiechowicz, Jakub and H{\"a}nggi, Peter and {\L}uczka, Jerzy},
  journal={Entropy},
  volume={24},
  number={1},
  pages={98},
  year={2022},
  publisher={MDPI}
}

@article{spiechowicz2013absolute,
  title={Absolute negative mobility induced by white Poissonian noise},
  author={Spiechowicz, Jakub and {\L}uczka, Jerzy and H{\"a}nggi, Peter},
  journal={Journal of Statistical Mechanics: Theory and Experiment},
  volume={2013},
  number={02},
  pages={P02044},
  year={2013},
  publisher={IOP Publishing and SISSA}
}

@article{marchenko2025approach,
  title={Approach to nonequilibrium: From anomalous to Brownian diffusion via non-Gaussianity},
  author={Marchenko, Ivan G. and Marchenko, Igor I. and {\L}uczka, Jerzy and Spiechowicz, Jakub},
  journal={Chaos: An Interdisciplinary Journal of Nonlinear Science},
  volume={35},
  number={2},
  year={2025},
  publisher={AIP Publishing}
}

\addtocontents{toc}{\protect\newpage}
\phantomsection

\newpage
\thispagestyle{empty}
    \addcontentsline{toc}{section}{Scientific Articles}

\begin{center}
\LARGE
\hspace{0pt}
\vspace{0.5cm}
\vfill
SCIENTIFIC ARTICLES
\vfill
\hspace{0pt}
\end{center}

\end{document}